\documentclass{article}

\usepackage{arxiv}

\usepackage{textpos}
\usepackage[numbers]{natbib} % for bibliography
\usepackage{tabularx,longtable,booktabs,multirow,subfig,caption}%hangcaption
\usepackage{url} % URLs
\usepackage[english]{babel}
\usepackage{amsmath}
\usepackage{graphicx}
\usepackage{dsfont}
\usepackage{epstopdf} % automatically replace .eps with .pdf in graphics
\usepackage{backref} % needed for citations
\usepackage{array}
\usepackage{latexsym}
\usepackage{bm} % vector
\usepackage[pdftex,pagebackref,hypertexnames=false,colorlinks]{hyperref} % provide links in pdf
\usepackage{algorithm,algpseudocode}
\usepackage{mathtools}
\usepackage{setspace}
\usepackage[edges]{forest}

\hypersetup{pdftitle={},
  pdfsubject={}, 
  pdfauthor={},
  pdfkeywords={}, 
  pdfstartview=FitH,
  pdfpagemode={UseOutlines},% None, FullScreen, UseOutlines
  bookmarksnumbered=true, bookmarksopen=true, colorlinks,
    citecolor=black,%
    filecolor=black,%
    linkcolor=black,%
    urlcolor=black}

\usepackage[all]{hypcap}

\def\@makechapterhead#1{%
  \vspace*{10\p@}%
  {\parindent \z@ \raggedright \sffamily
    \interlinepenalty\@M
    \Huge\bfseries \thechapter \space\space #1\par\nobreak
    \vskip 30\p@
  }}

\def\@makeschapterhead#1{%
  \vspace*{10\p@}%
  {\parindent \z@ \raggedright
    \sffamily
    \interlinepenalty\@M
    \Huge \bfseries  #1\par\nobreak
    \vskip 30\p@
  }}

\allowdisplaybreaks

\date{11 September 2023}

\title{Real-time VaR Calculations for Crypto Derivatives in kdb+/q}

\author{ \href{https://orcid.org/0009-0004-3971-7081}{\includegraphics[scale=0.06]{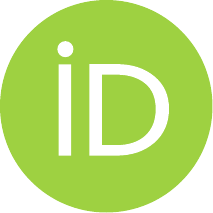}\hspace{1mm}Yutong Chen}\\
	Department of Computing\\
	Imperial College London\\
	South Kensington Campus\\
    London SW7 2AZ\\
    United Kingdom\\
	\texttt{ytchen102@outlook.com}\\
	\And
	\href{https://orcid.org/0000-0001-6846-6649}{\includegraphics[scale=0.06]{orcid.pdf}\hspace{1mm}Paul Bilokon}\\
	Department of Computing\\
	Imperial College London\\
	South Kensington Campus\\
    London SW7 2AZ\\
    United Kingdom\\
	\texttt{paul.bilokon@imperial.ac.uk}\\
    \And
	Conan Hales\\
	KX Systems\\
	Cannon Green Building\\
	27 Bush Lane\\
    London EC4R 0AN\\
    United Kingdom\\
	\texttt{chales@kx.com}\\
    \And
	Laura Kerr\\
	KX Systems\\
    Ormeau Avenue\\
    Co Antrim BT2 8HD\\
    N. Ireland\\
	\texttt{lkerr@kx.com}\\
}

\renewcommand{\shorttitle}{Real-time VaR Calculations}

\hypersetup{
pdftitle={Real-time VaR Calculations for Crypto Derivatives in kdb+/q},
pdfauthor={Yutong Chen, Paul Bilokon, Conan Hales, Laura Kerr},
pdfkeywords={risk, financial risk, real-time risk, value-at-risk, cryptocurrency, high-frequency data, big data, time series databases},
}

\begin{document}
\maketitle

\begin{abstract}
Cryptocurrency market is known for exhibiting significantly higher volatility than traditional asset classes. Efficient and adequate risk calculation is vital for managing risk exposures in such market environments where extreme price fluctuations occur in short timeframes. The objective of this thesis is to build a real-time computation workflow that provides VaR estimates for non-linear portfolios of cryptocurrency derivatives. Many researchers have examined the predictive capabilities of time-series models within the context of cryptocurrencies. In this work, we applied three commonly used models - EMWA, GARCH and HAR - to capture and forecast volatility dynamics, in conjunction with delta-gamma-theta approach and Cornish-Fisher expansion to crypto derivatives, examining their performance from the perspectives of calculation efficiency and accuracy. We present a calculation workflow which harnesses the information embedded in high-frequency market data and the computation simplicity inherent in analytical estimation procedures. This workflow yields reasonably robust VaR estimates with calculation latencies on the order of milliseconds. 
\end{abstract}

%%%%%%%%%%%%%%%%%%%%%%%%%%%%%%%%%%%%
\section{Introduction}
Since the creation of Bitcoin over a decade ago, cryptocurrencies have undergone a notable transformation, evolving from a speculative concept to a distinct asset class known as virtual assets. These digital assets are now increasingly acknowledged by investors as a diversifier for their portfolios~\cite{FERKO2023100778}. The growth of derivative markets has contributed to enhanced market efficiency and liquidity, yet it has also intensified price jumps and mass liquidation events, as observed in the market of Bitcoin~(BTC)~\cite{ALEXANDER2023478, jalan2021effect} and Ethereum~(ETH)~\cite{bitmex}. The significant fluctuations in the underlying cryptocurrency assets, amplified through the leverage of derivative products, expose market participants to exceptionally high risks. Consequently, investors need adequate tools for making informed investment choices and managing market risk.

In this thesis, we focus on the application of \emph{Value-at-Risk~(VaR)}, which is the most prominently adopted risk management tool for assessing market risk, to portfolios consisting of cryptocurrency derivatives, specifically the futures and options traded on Deribit. By focusing on downside risk, VaR answers the fundamental question from investors: how much loss might my portfolio incur over some investment horizon. 

While employing a sophisticated inference model for capturing volatility dynamics, together with a simulation-based approach for VaR estimation, often leads to a better comprehension of risk exposures, its practicality can be hindered when considering the cost of latency which arises from the use of stale market data. It is necessary to balance the trade-off between accuracy and latency, especially in the context of a high-frequency trading environment.

This thesis aims to develop a real-time solution for VaR calculation tailored to cryptocurrency derivative portfolios. This solution aims to provide reasonably accurate risk estimation while maintaining computational efficiency. To accomplish this objective, this thesis leverages the capabilities of kdb+, a specialised database designed to support real-time analytics on high-frequency time-series data.

\subsection{kdb+ and q}
kdb+ is a vector-oriented in-memory database developed by KX~\cite{kx}. The database is optimised for real-time analysis on large and continuously growing volumes of time-series data, making it a popular choice among financial institutions. q is the built-in array programming and query language in kdb+.

The outstanding performance by kdb+ comes from its optimised data storage workflow, columnar representation of data and its small footprint measuring just over 800kb~\cite{kdbspeed}. To support immediate query on ingested data, data is first published to real-time database~(RDB) in memory. At the end of each day, the data is then migrated to an on-disk historical database~(HDB) where it is stored as memory mapped files, eliminating CPU operations required for object translation between memory and disk. The columnar format allows for efficient disk writes and facilitates more targeted data retrieval when queried with q-SQL, which mimics SQL syntax for convenience, but operate on tables via vector operations on column lists, rather than a row-by-row basis.

The kdb+ implemented a unique tick capture architecture, kdb+tick~\cite{kdbtick}, which subscribes to incoming high-frequency data, and updates RDB or performs relevant analytics in real time as customised by q scripts. The details on the specific use case in this thesis will be discussed later.

\begin{figure}[ht]
\centering
\includegraphics[width = 0.9\hsize]{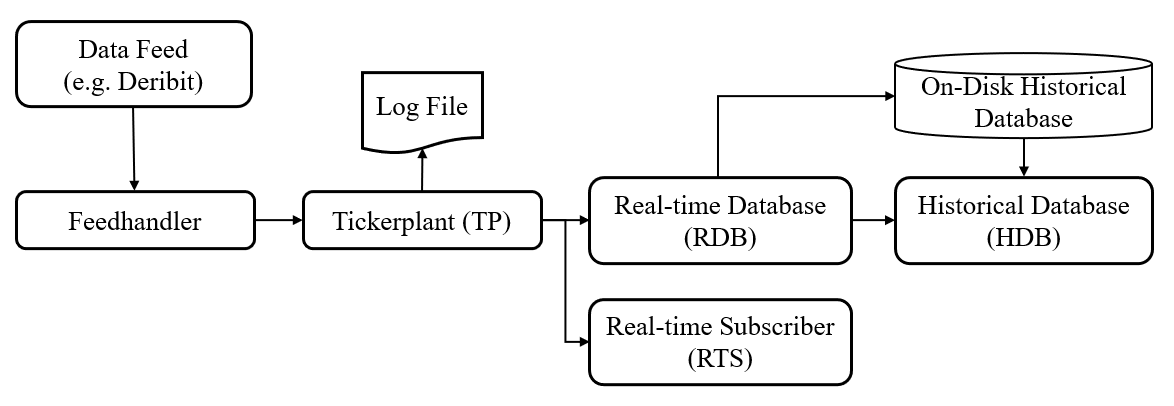}
\caption{kdb+tick architecture~\cite{kdbtick} }
\label{fig:kdbtick}
\end{figure}

\subsection{Deribit}
In this thesis, financial time series data are sourced from Deribit exchange, a leading cryptocurrency exchange that specialises in the futures and European-style options market. Launched in June 2016, Deribit now holds over 90\% of the market share in crypto options trading, with average daily trading volume of over 1 billion dollar~\cite{deribit}. Notably, Deribit offers both futures and European style cash-settled options for Bitcoin~(BTC) and Ethereum~(ETH), which are crypto derivatives examined in this thesis. Therefore, utilizing data from Deribit ensures adequate coverage of the relevant financial instruments of interest.

\subsection{Contribution}
Recognising the importance of appropriate risk management for investors participating in the cryptocurrency derivatives market, this work aims to develop a system that calculates Value-at-Risk~(VaR) for cryptocurrency derivatives portfolios in real time. The main contributions of our work are as follows:

\begin{itemize}
    \item \textbf{design of a real-time VaR computation workflow}: we developed a step-by-step procedure that applies analytical approaches at each stage of the computation. The workflow starts with applying a parsimonious volatility model in conjunction with efficient OLS estimators to high-frequency market data for volatility forecasting. Subsequently, the parametric approaches of delta-gamma-theta approximation and Cornish-Fisher expansion are employed to perform VaR estimation for crypto portfolios. Replacing the commonly used simulation approach with a reasonably robust analytical approach, our architecture produces real-time VaR forecast based on the most recent market data available.   
    
    \item \textbf{implementation of a VaR calculation system in kdb+ and q}: we implemented a kdb+ tick architecture and a VaR calculation system in q. The tick architecture was customised to pre-process the high-frequency market data, thus facilitating more efficient and robust calculation process. The calculation system was developed to leverage the in-memory compute engine to optimise the calculation latency. When tested on a portfolio holding all available cryptocurrency derivatives on Deribit, our architecture successfully delivered VaR estimations with the shortest latency at 14.2 milliseconds and the space usage of 1MB.
    
    \item \textbf{implementation of a web-based interface in KX dashboards}: to enhance the accessibility of the calculation system, we also developed a web-based interface to deliver calculation results and other associated market metrics.
    
    \item \textbf{a comparative analysis of three common volatility models}: as part of the work, we examined the trade-off between calculation latency and accuracy for using EWMA, DCC-GARCH and HAR models in real-time volatility forecasting, in the context of tail risk metric Value-at-Risk~(VaR). 
\end{itemize}

\subsection{Structure}

The remainder of this work is structured as follows: 

\textbf{Section~\ref{chap:Background}} begins with an overview of fundamental financial concepts and instruments relevant to this thesis. We then detailed the common approaches in calculating the VaR metric. Furthermore, we reviewed prior researches by others on the stylised facts in cryptocurrency market and the popular volatility models used to capture these observed dynamics.

\textbf{Section~\ref{chap:DesignImple}} presents the design principles and implementation details of the real-time VaR calculation system. The system comprises three key components: data service, calculation service and user inference. Following this sequence, we thoroughly 
 examined the design and implementation choices made.

\textbf{Section~\ref{chap:Evaluation}} elaborates on the evaluation of the computation workflow. This evaluation is conducted from two dimensions: calculation latency and VaR estimate accuracy. We analysed and summarised the results derived from these empirical assessments.

\textbf{Section~\ref{chap:Conclusion}} concludes on this work summarising the achievements of this work and discussing potential areas for future work.

%%%%%%%%%%%%%%%%%%%%%%%%%%%%%%%%%%%%
\section{Background}\label{chap:Background}
\subsection{Derivatives}
Derivatives are financial products that derive values from the performance of a specific or basket of financial instruments referred to as \emph{underlyings}. Derivatives are typically represented as contracts between two or more parties, specifying terms and conditions under which payments would be made from one party to the other. Use cases of derivatives include hedging, speculation and leveraging. Additionally, derivatives also provide an alternative medium to gain exposure to specific assets, offering flexibility and potentially lower costs compared to direct ownership. Common derivative products include options, futures, forwards and swaps. The former two are introduced in the following sections as they are the focus of this thesis.

\subsubsection{Futures}
Futures are standardised exchange-traded contracts that imply an obligation for the parties involved to buy or sell the underlying asset at a predetermined price, known as the \emph{strike price}, and on a specified future date, referred to as the \emph{expiration date} or \emph{maturity date}. The payoff for futures buyer at expiration is calculated as the difference between price of underlying asset $S_t$ and strike price $K$, implying a linear relationship between $S_t$ and $K$ as shown in Figure~\ref{subfig:futures}.

\subsubsection{Options}
In contrast to futures contracts, options represent the right, rather than the obligation, to buy or sell the underlying asset at a predetermined strike price and on or before a specified expiration date. Call options grant the holder the right to buy, and put options grant the holder the right to sell. Depending on when the right to buy or sell can be exercised, options are categorised as European-style or American-style. European-style options allow the holder to exercise the right on the expiration date, while American-style options provide the holder with the flexibility to exercise the right at any time before the expiration date.

\begin{figure} \centering
     \subfloat[futures]{
        \centering
         \includegraphics[width=0.3\textwidth]{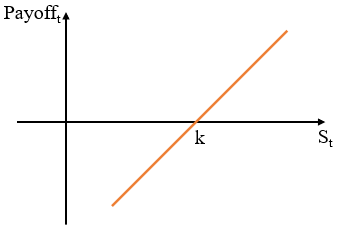}
         \label{subfig:futures}
    }
     \subfloat[call options]{
     \centering
         \includegraphics[width=0.3\textwidth]{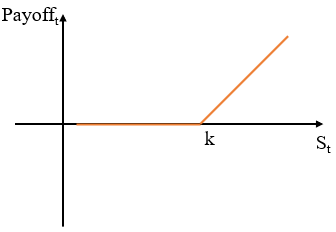}
         \label{subfig:call}
    }
     \subfloat[put options]{
     \centering
         \includegraphics[width=0.3\textwidth]{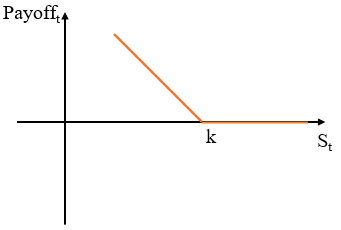}
         \label{subfig:put}
    }
        
        \caption{Payoff of futures and options at expiration for derivative buyer/holder.}
        \label{fig:payoff}
\end{figure}

One important implication of this right to buy or sell is that when the underlying price is not in favour of the option holder, the holder could choose not to exercise it. This discretion introduces non-linearity into the relationship between the movement of the underlying asset price and the change in corresponding options positions, where payoff for option holders is floored at $0$, as illustrated in Figure~\ref{fig:payoff}.

\subsubsection{Cryptocurrency Derivatives}
A important feature that distinguishes cryptocurrency derivatives from derivatives on traditional assets is the mechanism of cash settlement. In traditional derivatives, cash settlement will takes place in quoted currency, which is the currency used to price the underlying.

To illustrate, in the case of a futures contract on the British Pound with a strike price denominated in US Dollars, the cash payment would be made in US Dollars. Conversely, for majority of cryptocurrency derivatives, particularly those traded on Deribit, profits or losses will be settled in the corresponding cryptocurrency, akin to the British Pound in the aforementioned example.

\subsection{Value-at-Risk~(VaR)}
\begin{figure}[ht]
\centering
\includegraphics[width = 0.5\hsize]{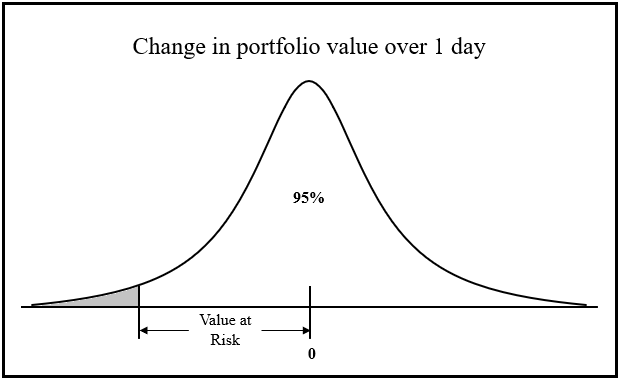}
\caption{95\% 1-day VaR for a hypothetical portfolio representing the 5 percentile of the distribution for change in portfolio value over 1 day. }
\label{fig:var}
\end{figure}

Formally introduced with the RiskMetrics Technical Documents by JP Morgan in 1996, Value-at-Risk~(VaR) has evolved to an established measure of market risk exposure in the banking industry since the Basel Accord III mandated the use of VaR~\cite{SHARMA2012234}. For a given portfolio, VaR measures the maximum potential loss in market value that is not expected to be exceeded over a given time horizon $t$ with a probability of $p$, assuming no trading of portfolio positions occurred during $t$~\cite{holton2013var}. More formally, let $R_t$ be a random variable representing change in portfolio value over time horizon $t$, VaR represents the $1-p$th quantile in the distribution of $R_t$, such that
 \begin{equation}
     P(R_t\le VaR) = 1-p
 \end{equation}
 While the choice of $p$ and $t$ are discretionary, common parameters are 1-day and 2-week for $t$ and 95\% and 99\% for $p$. Figure~\ref{fig:var} provides an example of 95\% 1-day VaR.

\subsubsection{Measuring VaR}
According to Holton~\cite{holton2013var}, VaR is a risk measure that combines exposure and uncertainty in its representation. Therefore, VaR estimation typically involves 3 procedures: 
\begin{itemize}
    \item a mapping procedure to quantify exposure: \par
    By considering a vector of market variables $\bm{R_t}$, this procedure is to express the values of the portfolio \begin{math}{P_t}\end{math} as a function of these variables. For instance, in the case of a portfolio holding Bitcoin call options with strike $k$ and expiration date $t$, portfolio value can be expressed as \begin{math}P_t = f(\bm{R_t}) = \max \{\bm{R_t-K},\bm{0}\}\end{math}, where \begin{math}\bm{R_t}\end{math} represents is the market variable vector containing only Bitcoin price. 
    \item a inference procedure to quantify uncertainty: 
    
    To quantify uncertainty in market variables over time $t$, the procedure starts with assuming them as random variables following a stochastic process and model their movements via some choice of distributions, such as log-normal or Student $t$, thus arriving at their conditional distribution at time $t$ based on information available at time $t-1$. The widely applied EWMA and GARCH model will be discussed in Sections~\ref{EWMA} and \ref{GARCH}.
    \item a transformation procedure to combine exposure and uncertainty: 
    
    Leveraging the result from previous two procedures, risk is presented with a characterisation of conditional distribution of portfolio value, in alignment with the quantile concept of VaR. Various approaches exist to compute or estimate this risk measure. In general, higher accuracy comes at the cost of computation complexity. In the next section, we present the two main classes of these approaches.
\end{itemize}
\begin{figure}[ht]
\centering
\includegraphics[width = 0.5\hsize]{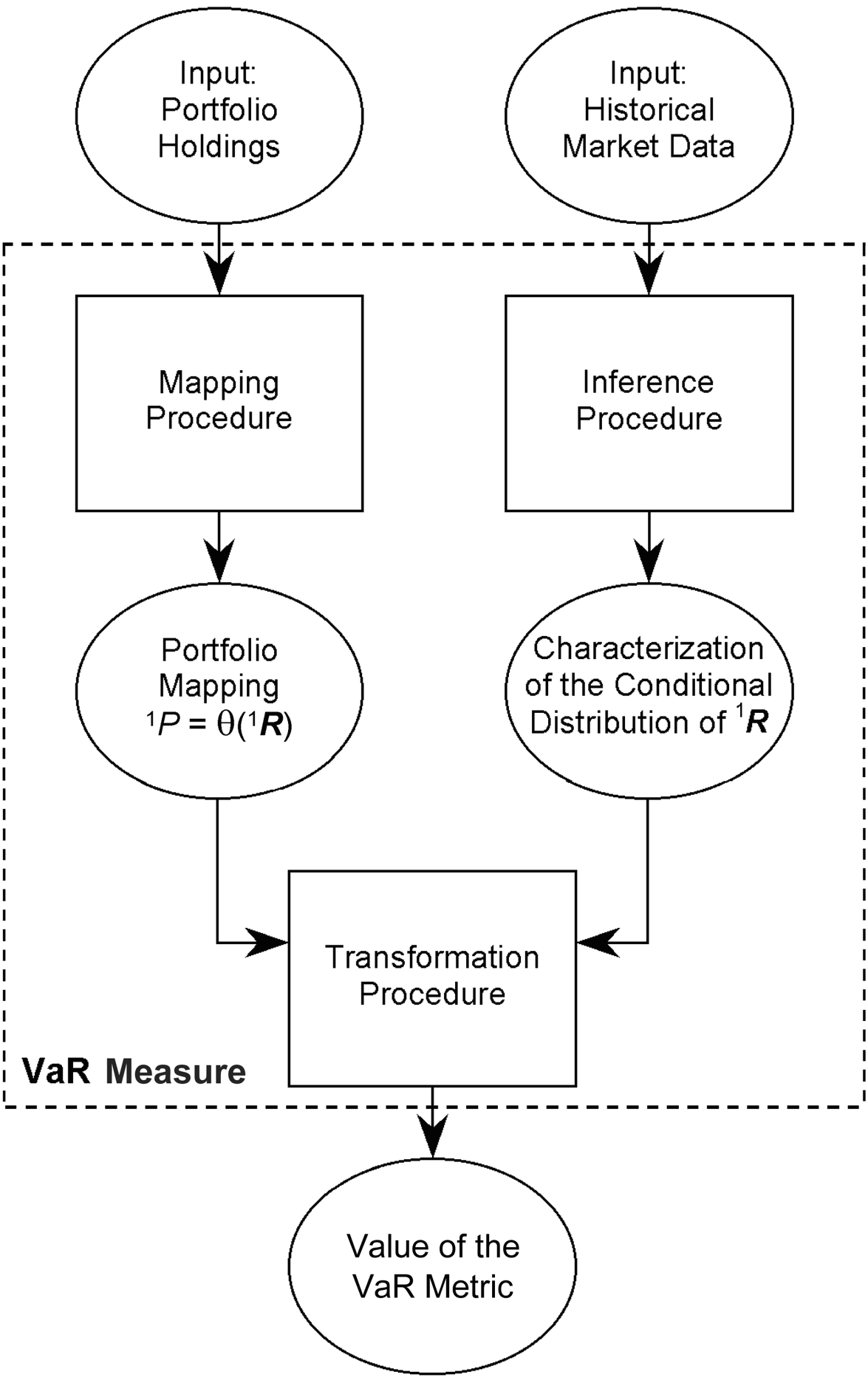}
\caption{Methodology for measuring VaR~\cite{holton2013var}. }
\label{fig:proc}
\end{figure}
 
\subsubsection{Parametric Approach}
Parametric approach, also known as variance-covariance approach, transforms conditional distribution of market variables to that of portfolio with analytical equations. A simplistic example is to assume joint Gaussian distribution on market variables \begin{math}\bm{R_t} \sim \mathcal{N}(\mu_t,\,\sum_t)\end{math}, where \begin{math}\mu_t\end{math} and \begin{math}\sum_t\end{math} are estimated from historical market data. If portfolio can be expressed as a linear transformation of these Gaussian random variables through \begin{math}P_t = f(\bm{R_t}) = \bm{a}\bm{R_t} + b\end{math}, portfolio value would also follow Gaussian distribution, \begin{math}P_t \sim \mathcal{N}(\bm{a}\mu_t+b,\,\bm{a^T}\sum_t\bm{a})\end{math}. With a tractable form, portfolio VaR can be derived analytically as a defined number of standard deviations below expectation.

\subsubsection{Non-Parametric Approach}
Non-parametric approach relies on simulation for transformation. Changes in market variables are sampled and the portfolio is re-evaluated on each sampled market conditions. The estimation of VaR is then obtained by applying an appropriate sample estimator to these re-evaluations.

Depending on the specific sampling procedure employed, transformation can be further categorised into Monte Carlo transformation or historical transformation: in the former, changes in market variables are obtained from applying Monte Carlo method, which involves simulating scenarios based on specified distributions; in the latter, changes in market variables are sampled from historical data, thereby capturing the observed variability in the past market conditions.

\subsection{Volatility Model}
With the widely used assumption that conditional mean of return equals zero, and the argument that construction of conditional mean is largely derived from economic theories on behaviours of market variables~\cite{holton2013var}, most of the work in the inference procedure is to model the conditional variance and covariance of market variables. This section first introduces some stylised facts observed in cryptocurrencies' time series data, then brings out the common models used in volatility forecast.

\subsubsection{Dynamics in cryptocurrency returns}
When assessing price movement in financial instruments, it is commonly assumed that the focus is to model the log returns observed in the time series data via
\begin{equation}
    r_t = \ln{\frac{P_t}{P_{t-1}}}.
\end{equation}
Firstly, returns are more comparable than prices. To illustrate, BTC trades around USD 27000 while ETH trades around USD 1900, a price movement of USD 1000 implies a mild change in BTC market but a significant one in ETH. Secondly, additive property of log returns linearises the effect of compounding. When analysing cumulative returns over multiple periods, log returns can be summed directly. Thirdly, the non-negativity of the argument of log function is well aligned with such property in any asset price.

Below are some stylised facts consistently found in existing literature on cryptocurrency:
\begin{itemize}
  \item \textbf{Volatility Clustering:} empirical analysis~\cite{FUNG2022102544, BAUR2018148} established statistically significant evidence for conditional heteroskedasticity in returns behaviour, specifically the pattern that large change in crypto prices tend to be followed by large changes. Volatility shows a tendency to persist, and in statistical term, to auto-correlate. This is one of the key features leveraged in volatility forecast.
  \item \textbf{Volatility Asymmetry:} studies examining different periods of cryptocurrency market have found consistent evidence that volatility responds asymmetrically to past returns~\cite{BAUR2018148, CHEIKH2020101293, PANAGIOTIDIS2022101724, CHI2021101421}. For majority of cryptos, researchers observed reverse leverage effect that positive shocks increasing volatility more than negative shocks, a feature that resonates with safe-haven assets, such as gold. For BTC and ETH, however, such safe-haven feature has diminished over time. In studies covering more recent data by Gupta et al~\cite{jrfm15110513} and Aharon et al~\cite{AHARON2023102651}, it is noted that following the introduction of derivatives and expansion in market capitalisation, BTC and ETH resemble stocks where leverage effect dominates. 
  \item \textbf{Excess Kurtosis:} consistently large excess kurtosis was observed for major cryptocurrencies~\cite{BAUR2018148,TAN2021101377}, implying that their return distributions have heavier tails than Gaussian distribution. In particular, in Fung's comprehensive analysis covering 254 cryptos, it was observed that the kurtosis ranges from 4.63 up to 283.61~\cite{FUNG2022102544}. 
  
  \item \textbf{Strong Intra-market Correlation:} evidence of volatility co-movement has been identified among pairs of major cryptoassets, but as illustrated in Figure~\ref{fig:corr} such the direction of conditional correlation is unstable~\cite{KATSIAMPA2019221}. This may explain the poorer out-of-sample forecasting power of multi-variate GARCH model observed in Chi et al's work~\cite{CHI2021101421}, where the model including interactive effects between BTC and ETH had lower adjusted R-square than the univariate GARCH model.
  \begin{figure}[htb]
    \centering
    \includegraphics[width = 0.5\hsize]{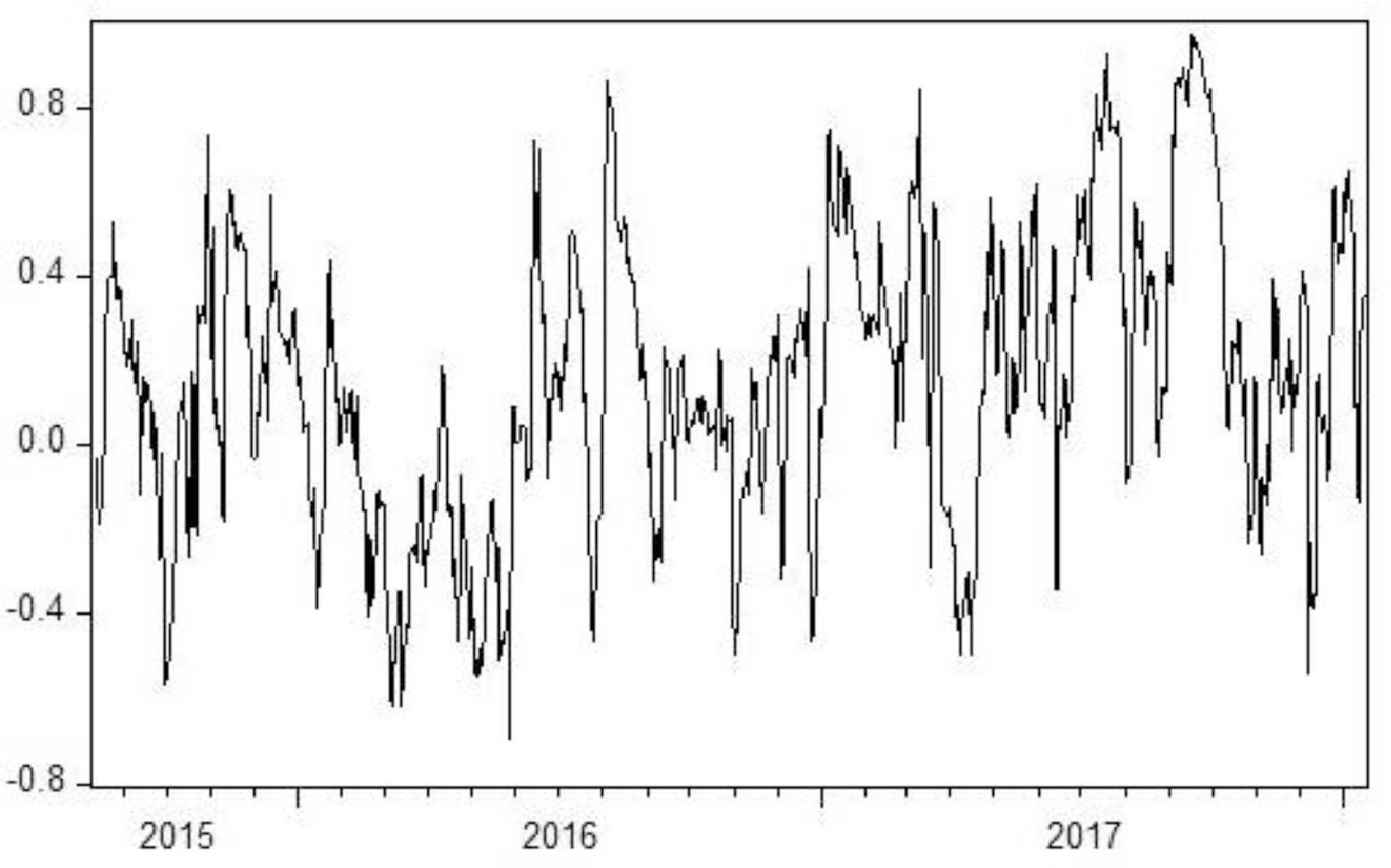}
    \caption{Conditional correlation between the price returns of Bitcoin and Ethererum ranging from -0.70 to 0.96~\cite{KATSIAMPA2019221}.}
    \label{fig:corr}
    \end{figure}
  \item \textbf{Structural Break:} due to exogeneous factors, such as geopolitical factors, social events, sudden changes in unconditional variance of asset returns occur, implying structural breaks in the volatility model. While researches found evidence of structural breaks in cryptocurrency markets via relevant statistical tests, they also noted that failure to account for structural breaks lead to overestimation of volatility persistence~\cite{ AHARON2023102651,ABAKAH2020680}.

\end{itemize}

\subsubsection{EWMA model}\label{EWMA}
Implemented as part of RiskMetrics forecasting methodology~\cite{RiskMetric}, Exponentially Weighted Moving Average (EWMA) is a simple but powerful measure to capture the dynamics in volatility and covariance. Since volatility reacts quickly to shocks, and the effect of shock declines exponentially as time passed, EWMA estimator incorporates such mechanism by applying higher weights to more recent observations and lower weights to those further away, through a decay factor $\lambda$:
\begin{equation}\sigma_{T+1|1,2,...T}^2 = (1-\lambda) \sum_{t=1}^{T} \lambda^{t-1}(r_t-\Bar{r})^2 \end{equation} 
where \begin{math}0 < \lambda < 1\end{math}. The estimator can also be expressed in a recursive format for forecasting volatility: 
\begin{equation}\label{eq:ewma1}\sigma_{t|t-1}^2 = \lambda \sigma_{t-1|t-2}^2 + (1-\lambda) r_{t-1}^2 \end{equation}

For covariance forecast, a similar estimator can be applied to each pair of market variables:
\begin{equation}\sigma_{ij}^2 = (1-\lambda) \sum_{t=1}^{T} \lambda^{t-1}(r_{it}-\Bar{r}_i)(r_{jt}-\Bar{r}_j) \end{equation} 
 with an equivalent recursive form.
\begin{equation}\label{eq:ewma2}\sigma_{ij, t|t-1}^2 = \lambda \sigma_{ij,t-1|t-2}^2 + (1-\lambda) r_{it}r_{jt} \end{equation}

\subsubsection{GARCH model}\label{GARCH}
The consensual model to capture the volatility pattern of financial time series is the generalised autoregressive conditional heteroskedasticity~(GARCH) model proposed by Bollerslev~\cite{BOLLERSLEV1986307} and its variants. In the standard GARCH($p$, $q$) model, the conditional variance $\sigma_t^2$ is a weighted average of a constant, conditional variance in previous $p$ periods and error term which is also referred as innovation in previous $q$ periods.
\paragraph{GARCH(p,q)}
\begin{equation}r_t = \mu_t + \varepsilon_t\end{equation}
\begin{equation}\varepsilon_t = \sigma_tz_t,  z_t \sim \mathcal{N}(0,\,1)\end{equation}
\begin{equation}\sigma_t^2 = \omega + \sum_{i=1}^p \alpha_i \epsilon_{t-i}^2 + \sum_{j=1}^q \beta_j \sigma_{t-j}^2\end{equation}
where \begin{math}\omega>0, \alpha_i\ge0\end{math} and \begin{math}\beta_j\ge0\end{math}. The parsimonimious model used in most empirical study for crypto volatility is GARCH(1, 1) where $p$=1 and $q$=1. 

\paragraph{asymmetric GARCH}
Such variants of GARCH model are particularly popular in volatility forecast for cryptocurrency due to the well-recognised phenomenon of asymmetric behaviour. Common asymmetric GARCH models are presented in Table~\ref{tab:garchs}. 

The relative performance among these asymmetric models is dataset dependent. While GJR-GARCH showed better performance in terms of log likelihood than EGARCH in the data from August 2015 to December 2018 analysed by Cheikh et al~\cite{CHEIKH2020101293}, in the study by Fung et al~\cite{FUNG2022102544} using March 2019 to March 2021 data, TGARCH showed better results. 

\begin{table}[htb]
    \centering
    \caption{List of asymmetric GARCH models.}
    \begin{tabular}{ccc}
    \toprule
    Model & Volatility Equation \\
    \midrule
       EGARCH~\cite{nelson1991conditional} & 
       \begin{math}\log(\sigma_t^2) = \omega + \alpha(|\varepsilon_{t-1}| - E(|\varepsilon_{t-1}|)) + \gamma\varepsilon_{t-1} + \beta\log(\sigma_{t-1}^2)\end{math}\\
       GJR-GARCH~\cite{glosten1993relation} &
       \begin{math}\sigma_t^2 = \omega + \alpha\varepsilon_{t-1}^2 + \gamma I(\varepsilon_{t-1} < 0)\varepsilon_{t-1}^2 + \beta \sigma_{t-1}^2\end{math}\\
       TGARCH~\cite{zakoian1994threshold} &
       \begin{math}
       \sigma_t = \omega + \alpha|\varepsilon_{t-1}| + \gamma|\varepsilon_{t-1}|I(\varepsilon_{t-1} <0) + \beta \sigma_{t-1}\end{math}\\
    \bottomrule
    \end{tabular}
    \label{tab:garchs}
\end{table}

\paragraph{error term in GARCH model}
Despite GARCH with Gaussian error term presents a heavy-tailed behaviour in comparison to the Gaussian distribution, it often fails to incorporate such observed style in returns sufficiently~\cite{doi:10.1080/713665670}. Therefore, it is common to assume error terms follow non-Gaussian distributions, such as Student-t, skewed Student-t and generalised normal distribution. \par
In particular, the cross-sectional analysis by Fung et al~\cite{FUNG2022102544} concluded that Student's t error distribution in general have a better performance in describing dynamics in cryptoassets, followed by its skewed version. This is also consistent with the observation by Peng et al~\cite{peng2018worlds}that heavy-tailed distribution produces better results than Gaussian distribution.

\paragraph{HAR model}\label{HAR}
With the availability of high-frequency market data, a popular alternative to GARCH-type
estimators is the heterogeneous autoregressive~(HAR) model proposed by 
Corsi~\cite{Corsi2009174}, which utilises realised variance~(RV) as a non-parametric estimator to model latent quadratic variance. 

\paragraph{from Quadratic Variance to Realised Variance}
Let the logarithmic price $P$ of an asset follows a stochatic process defined as:
\begin{equation}
    dP_t = \mu(t)dt+\sigma(t)dW_t
\end{equation},
where $\mu(t)$ and $\sigma(t)$ are the corresponding drift and volatility processes respectively and $W_t$ is a standard Brownian motion~\cite{HAR}. Since the logrithmic transformation linearises the compounding effect in cumulative return, the return over time interval from $t$ to $t+\tau$ is:
\begin{equation}
    r_{t,\tau} = \int_{t}^{t+\tau} \mu(\tau)d\tau + \int_{t}^{t+\tau} \sigma(\tau)d\tau
\end{equation}
and the corresponding quadratic variance~(QV) is 
\begin{equation}
    QV{t,\tau} = \int_{t}^{t+\tau}\sigma^2(\tau)d\tau.
\end{equation}

Though $QV{t,\tau}$ cannot be observed directly, realised variance which can be treated as an observed variable in the presence of high frequency data is a consistent estimator of QV, as shown in the seminal work by Anderson and Bolleslev~\cite{26b2223b-2816-35cd-8184-da686b05fa62}. Realised variance refers the sum
of squared intraday returns over a defined time interval, with returns calculated at finer intervals~\cite{HAR}. For example, to calculate the RV over 1 day, we can divide the daily trading hours into 144 periods of 10-min and measure the return over each of them, thus RV is the sum of the 144 squared returns.

\paragraph{HAR-RV}
The standard HAR model presented in Corsi's work specified that the 1-step-ahead daily $RV_{t+h}$ can be modelled by HAR(3)-RV as:
\begin{equation}
    logRV_{t+1d} = \beta_0 + \beta_dlogRV_{t}^{(d)} + \beta_wlogRV_{t}^{(w)} + \beta_mlogRV_{t}^{(m)}+ \epsilon_t
\end{equation}

where $RV_{t}^{(d)}$, $RV_{t}^{(w)}$, $RV_{t}^{(m)}$ are the daily, weekly and monthly observed realised variance respectively~\cite{Corsi2009174}. Log of RVs is used to ensure the positiveness.

Though HAR-RV model is a simple model which can be fitted with Ordinary Least Squares~(OLS) estimators, it is capable of capturing the high persistence in most realised variance series and producing sound volatility forecast~\cite{Corsi2009174}. Previous work by Aalborg et al~\cite{AALBORG2019255} have found robust performance of HAR models in the context of Bitcoin volatility. In addition to these, Bergsli et al~\cite{BERGSLI2022101540} compared HAR models with GARCH models and reported superior performance of HAR models over any GARCH models as measured by mean squared error~(MSE) and such difference in performance is the largest for short-term forecasting horizons, such as 1 day.

Later studies have also explored other variants of the HAR model. For example, to correct for measurement errors in daily realised variance, Bollerslev et al proposed a model incorporating realized quarticity , which is referred to the HARQ model~\cite{BOLLERSLEV20161}. HARQ-type models improve the forecast performance by placing less weight on historical values of RV when the measurement error is higher. Furthermore, Yu also explored including leverage effect and jump components, and reported that leverage effect contributes more than jump components for out-of-sample volatility forecasting for BTC~\cite{YU2019120707}.

\paragraph{realised covariance estimator}
In parallel with the univariate semimartigale process of logarithmic price, the realised covariance estimator is derived from the covariation of the processes for two logarithmic prices. For example, Bollerslev et al have extended the univariate HARQ model to the multivariate space as the MHARQ model~\cite{BOLLERSLEV201871}.

Two issues arise as we move into multivariate realised measures: 1) asynchronous in transactions in different assets and 2) the Epps effect which describes the bias towards zero in realised correlation as the sampling frequency increases~\cite{Epps1979291}. Possible solutions to these issues include the work by Hayashi and Yoshida~\cite{hayashiyoshida}, which uses a sampling scheme include all overlapping intraday returns based on the actually observed price series, and the multivariate kernel estimator introduced by Barndorff-Nielsen et al~\cite{BARNDORFFNIELSEN2011149}.

%%%%%%%%%%%%%%%%%%%%%%%%%%%%%%%%%%%%
\section{Design \& Implementation}\label{chap:DesignImple}
This section details the design considerations and the associated implementation decisions of the system.

\subsection{Architecture Overview}
As shown in Fig~\ref{fig:architecture}, the processes in the computation workflow can be categorised into three groups:
\begin{figure}[htb]
\centering
\includegraphics[width = \hsize]{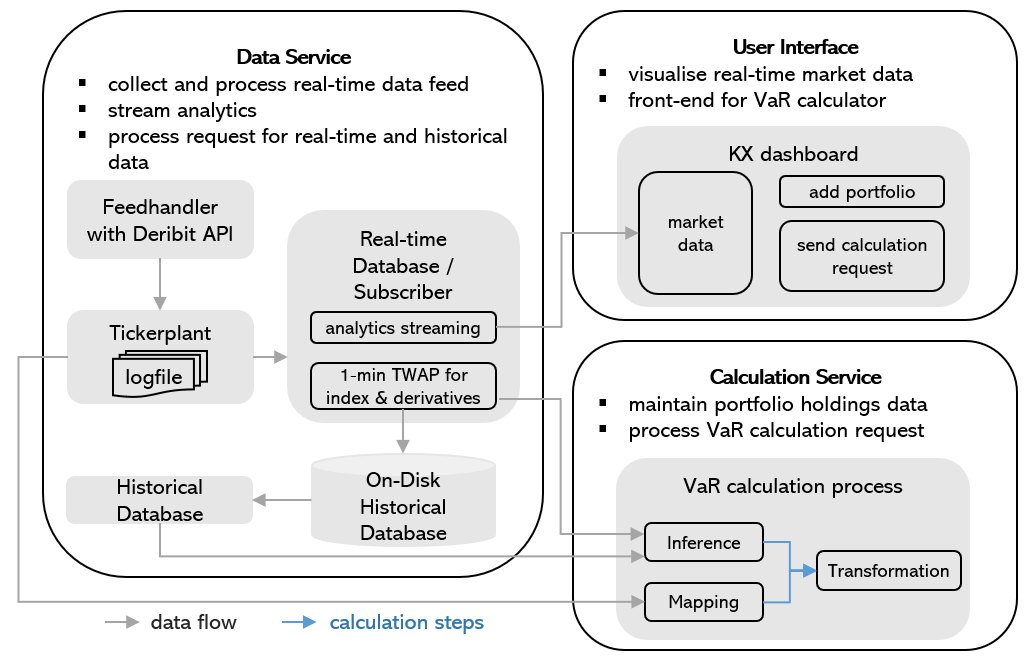}
\caption{Architecture Diagram}
\label{fig:architecture}
\end{figure}
\begin{itemize}
    \item data service responsible for collecting, processing and storing market data
    \item calculation service responsible for calculating VaR estimate in real time
    \item user interface responsible for delivering calculation result and other market data analytics
\end{itemize}

\subsection{Real-time Data Sourcing with kdb+}
Regarding the primary data source, we employed the Deribit API service, which supports the subscription to tick level data over Websockets. Tick data refers to timestamped entries of market data including bid and ask prices, last traded price, open interest, among other relevant metrics. Each transaction prompts the generation of a tick record associated with the traded product. For the period from 2023.04.11 to 2023.08.31, the average number of tick records received per day is $3.4 \times 10^7$. To handle such enormous volume of high-frequency data and create a real-time market data stream for Value at Risk~(VaR) calculation, we used the aforementioned kdb+ tick architecture in conjunction with this API service~\cite{kdbtick}. Below we introduce the components within this architecture.
  
\subsubsection{Feedhandler}
For Deribit data to be consumed by kdb+, we implemented a Python script as the feedhandler. It subscribes to tick-level data of all available futures and options products on Deribit as well as the price indices for BTC and ETH. The tick data are transmitted via a Websocket connection in JSON format. Upon receipt, the script maps the received data into the suitable table schema within tickerplant and then pushes them to the corresponding table in tickerplant for downstream distribution.

\subsubsection{Tickerplant}
As data are pushed in by the feedhandler, the tickerplant process acts as a publisher, distributing the received data to both real-time databases and subscribers. Simultaneously, it writes records to a log file on disk for data recovery purpose. Since we would like to use the latest market data in calculation workflow, we implemented the standard zero-latency tickerplant structure through which each update from the feedhandler gets written to disk and published to subscribers independently.

\subsubsection{Real-time Database and subscribers}\label{subscriber}
As complementary components of the publisher process, we implemented subscriber processes that subscribe to tables published by the tickerplant. To maintain a clear separation of responsibilities, we have implemented two distinct instances of real-time subscribers: namely, the price subscriber and the streaming subscriber.
    
\paragraph{Price subscriber} 
Price subscriber plays a pivotal role in supporting the core VaR calculation process. This subscriber is responsible for handling all queries on intraday market data relevant for VaR calculation. It achieves this by maintaining keyed tables that capture time-weighted market price of underlying cryptocurrency indices and individual products throughout the intraday trading period, namely \texttt{indextwap}, \texttt{futuretwap} and \texttt{optiontwap}. Keyed tables in kdb+ are dictionaries that, instead of mapping keys to values, link one table of keys to another table of values, such that each row in the keys table is mapped to a corresponding row in the values table\cite{kx}. The choice of keyed tables over simple tables is motivated by memory saving in using dictionary lookup and also to facilitate the \texttt{pj} used in the price subscriber.
    
For cryptocurrency indices table \texttt{indextwap}, prices used are the median of mid prices sourced from several major exchanges~\cite{deribitidx}. For futures and options products offered by Deribit, prices used are mark prices, which is a Deribit-specific metric representing the fair value of the specific products and being used in derivatives contracts valuation. 
    
The price subscriber pre-processes the incoming tick level data by aggregating it into the corresponding 1-min interval record before inserting it into the relevant table. The key reason for this preprocessing is to use the pre-averaged prices to mitigate the impact of market microstructure noise, a topic we will delve in further in Section~\ref{rv}. In addition, since the tick data for different indices and products arrive at random times, the averaging process synchronises market data. Furthermore, this significantly reduces the amount of intraday data held in memory.

\begin{algorithm}[H]
    \caption{Preprocess tick data to 1-min TWAP data}\label{alg:updIndex}
    \centering
    \begin{algorithmic}
        \small
        \Require $d$, \texttt{INDEXTWAP}
        \Function{updIndex}{$d$}
        \If{$d$'s type $=$ list} \Comment{$d$'s type is list if it's replayed from tickerplant logfile}
        \State $d \gets \textproc{listToTable}(d)$
       \EndIf
        \State $d \gets \textproc{plusjoin}(d,\texttt{INDEXTWAP})$
        \State $d \gets \textproc{updateAverage}(d,\texttt{INDEXTWAP})$
        \State $\textproc{upsertToTable}(d,\texttt{INDEXTWAP})$
        \EndFunction
        \end{algorithmic}
\end{algorithm}

In Algorithm~\ref{alg:updIndex} above we use pseudocode to illustrate the pre-processing function for published index tick data. Similar functions exist for published futures and options tick data. The update for TWAP data is delivered through the plusjoin and updateAverage steps. The plus-join \texttt{\symbol{92}pj} is a native semantic in kdb+ which left joins \texttt{INDEXTWAP} to $d$ and applies addition to duplicate columns. 

At the end-of-day, the price subscriber will unkey these tables and save them to Historical Database~(HDB) on disk using the built-in function \texttt{.Q.dpfts}. This archival process ensures their availability for subsequent inference procedures from the following trading day onwards.

\paragraph{Streaming subscriber} 

This subscriber process serves as a publisher of real-time market analytics designed for consumption by downstream dashboards. Additionally, it facilitates data queries originating from the KX dashboard. At 1-second intervals, this subscriber extracts implied volatility values for each option product categorised by maturity and strike. These values are then used to generate the volatility surface for individual underlying cryptocurrencies, subsequently publishing them to the downstream KX dashboard. Furthermore, as it receives tick records from the upstream publisher tickerplant, the subscriber maintains \texttt{future} and \texttt{option} tables of raw tick data to support additional analytics queries from the dashboard, such as open,low, high, close prices over specified intervals. To optimise memory usage, the streaming subscriber routinely purges earlier entries in its tables over the course of the day.

\subsubsection{Historical Database}
Historical market data are used in the inference procedure of VaR calculation as well as in the backtests performed in evaluation. They are accessed via Historical Database~(HDB) process. HDB process holds data before the current date with tables stored on disk~\cite{hdb}. 

For best efficiency for search and retrieval, data are stored partitioned by date and splayed by column, such that they are divided into separate partition directories named after the date. Inside each date directories are directories named after the table names, which contains separate files for each splayed column.

The partitioned structure limits operations to relevant columns. For example, when performing inference procedures detailed in Section~\ref{inference}, we only require columns of \texttt{sym}, \texttt{time} and \texttt{twap} from \texttt{INDEXTWAP} table, therefore the data query will deserialise into memory only files for the columns it requires.

\subsubsection{Interprocess Communication among kdb+ instances}
During real-time calculation, the calculation process needs to query data processes to obtain historical and real-time market data to perform calculation procedures. This is achieved by the built-in IPC functionality in q. Since in our implementation, the processes run on the same machine, Unix domain sockets are preferred over localhost IPC connections as it often has lower latency and higher throughput.

For a kdb+ process, such as real-time subscriber and HDB, to listen on a port, it is started with command line parameter \texttt{-p}. These allow data processes to wait on incoming connections. For the calculation process to communicate with data processes, we used \texttt{hopen} to open connections to real-time subscriber and HDB and obtain the respective connection handles from return values. As data query is required, we use the appropriate connection handle to message the data processes. In our implementation, these remote queries were wrapped in separate utility functions, such as \texttt{.util.getidxtwap} for obtaining twap data for inference procedures.

\subsection{Value-at-Risk Calculation}
To calculate VaR for a portfolio $P_i$, the system first queries historical market data from HDB process to perform inference on the conditional distribution of underlying cryptocurrencies, then uses real-time market data from market price subscriber to map and transform it to the conditional distribution of portfolio return. Below we provide an overview of the calculation process, followed by details on the implemented inference, mapping and transformation procedures.

\subsubsection{Calculation Workflow}\label{workflow}
The key inputs for portfolio VaR include portfolio holdings, confidence level and time horizon. We maintain a table \texttt{portfolio} in the calculation process to track holdings added to each portfolio which is identified by column \texttt{\textasciigrave pid}. Anticipating changes in portfolio holdings from time to time, it is essential to acknowledge that the duration of this holdings table is tied to the specific calculation process. This table only resides in the memory associated with the process and gets erased when the process terminates.

The workflow starts by identifying portfolio holdings and relevant cryptocurrency indices for which it needs to perform inference procedures. Derivative products on Deribit follow the naming convention of \texttt{crypto-maturity(-strike-optiontype)}. For example, for BTC future maturing on 2023.12.29, its identifier is \texttt{BTC-29DEC23}. This allows us to parse the corresponding underlying index from their identifiers. For the scope of this work, we only have derivatives on BTC and ETH. This implies at most two indices are relevant for the inference procedure. 

The process then communicates with data processes to query for historical market data. Depending on the inference algorithms chosen, the time series data for TWAP are processed into log return series or realised variance series, then get passed  to inference algorithms, together with VaR time horizon, to model and forecast the conditional distribution of underlying cryptocurrency indices. For data used for mapping procedures, they are maintained by keyed tables \texttt{LatestProduct} and \texttt{LatestIndex} residing in the calculation process. The process can query these tables directly for real-time market data on current price, greeks for options and pass them to mapping algorithms to establish a tractable representation of portfolio returns in terms of indices returns.

The output from these two algorithms, namely the forecast and the portfolio mapping are used as inputs in transformation algorithms to generate an estimation for the corresponding quantile of portfolio return as specified in confidence level. Finally, this quantile of return is transformed to absolute value, which is the VaR estimate.

We summarise the calculation workflow in Algorithm~\ref{alg:var} below.

\begin{algorithm}[htb]
    \caption{Portfolio VaR calculation}\label{alg:var}
    \centering
    \begin{algorithmic}
        \Require $pid$, $ci$, $t$, $\texttt{portfolio}$
        \Function{.VaR.estimate}{$d$}
        \State $p \gets \textproc{getPortfolioPositions}(pid)$
        \State $idx \gets \textproc{extractIndex}(p)$
        \State $idxdata \gets \textproc{getInferenceData}(idx)$
        \State $tickdata \gets \textproc{getRealTimeTick}(p)$
        \State $cov \gets \textproc{inference}(idxdata, t)$
        \State $greeks \gets \textproc{mapping}(p, tickdata, t)$
        \State $quantile \gets \textproc{transformation}(cov, greeks, ci)$
        \State \Return $quantile * marketValue$
        \EndFunction
        \end{algorithmic}
\end{algorithm}

\subsubsection{Inference Procedure}\label{inference}
The purpose of inference procedure is to characterise the distribution of factors motivating portfolio value changes, conditional on all information available as of time $t$. Within the scope of this work, underlying cryptocurrency prices, namely the \texttt{btc\_usd} and \texttt{eth\_usd} index prices on Deribit were employed as such factors. 

A widely used assumption in the inference procedure is that time series of cryptocurrency returns exhibit a conditional expectation of zero. While this assumption may be violated over longer time horizon, its applicability remains pertinent for the context of VaR calculation which often focuses on time horizon ranging from intraday to a maximum of 2 weeks. This assumption implies the key output of the inference procedure is a covariance matrix for crypto indices relevant to the portfolio holdings.

\paragraph{Analysis of tick data}\label{eda}
Before introducing the inference models, we start by examining the return series of Bitcoin~(BTC) and Ethereum~(ETH) to understand the statistical properties of these time series data and the associated challenges in modelling their volatility dynamics.The data used in this exploratory analysis were collected with the aforementioned kdb+ tick architecture. The dataset covers the period from 2023.04.11 to 2023.07.31\footnote[1]{Due to network issues, data were incomplete for certain days within the period. Since we use log returns over a fixed interval as samples, raw data have been processed such that missing data only reduced the total number of samples for analysis and did not distort the observed statistical properties.}.

The return series are calculated as the natural logarithmic differences of the TWAP of different averaging intervals as below:
\begin{equation}
    R_{it} = \ln{\frac{P_{i,t}}{P_{i,t-\tau}}}
\end{equation}
where 
\begin{align*}
    P_{i,t} &= \text{TWAP in USD of crypto i by averaging ticks received between ($t-\tau$,$t$]}\\
    \tau &= \text{interval at which TWAP is computed} 
\end{align*}

\begin{table}[htb]
\small
\centering
\caption{Descriptive statistics of BTC and ETH return series measured in USD. Vol(.p.a) represents volatility annualised by scaling raw standard deviation with $\sqrt{\frac{1440}{\text{no. of minutes in the interval}}*365}$.}
\begin{tabular}{llrrrrr}
\toprule
Crypto & Interval & Observations & Mean & Vol(.p.a) & Skewness & Kurtosis \\
\midrule
\multirow{10}[3]{*}{BTC} & 1-min & 102871 & 0.0000\% & 32.1\% & -1.83 & 75.75 \\
& 5-min & 20579 & 0.0002\% & 39.9\% & -1.34 & 48.7 \\
& 10-min & 10290 & 0.0004\% & 39.3\% & -1.42 & 43.4 \\
& 15-min & 6860 & 0.0005\% & 39.5\% & -0.40 & 30.0 \\
& 30-min & 3428 & 0.0011\% & 41.0\% & -1.03 & 34.2 \\
& 1-h & 1712 & 0.0019\% & 41.4\% & -1.98 & 38.8 \\
& 2-h & 857 & 0.0026\% & 39.4\% & -1.12 & 23.0 \\
& 6-h & 286 & 0.0072\% & 36.3\% & 0.16 & 7.4\\
& 12-h & 143 & 0.0185\% & 34.9\% & 0.47 & 5.2 \\
& 1-d & 73 & -0.0036\% & 37.7\% & $<$0.00 & 4.5 \\
\midrule
\multirow{10}[3]{*}{ETH} & 1-min & 102871 & -0.0001\% & 35.2\% & -0.51 & 138.9 \\
& 5-min & 20579 & -0.0003\% & 42.7\% & -0.35 & 59.6 \\
& 10-min & 10290 & -0.0006\% & 42.1\% & -0.63 & 49.4 \\
& 15-min & 6860 & -0.0009\% & 42.9\% & 0.03 & 42.2 \\
& 30-min & 3428 & -0.0018\% & 44.1\% & -0.88 & 34.3 \\
& 1-h & 1712 & -0.0039\% & 45.3\% & -2.02 & 37.9 \\
& 2-h & 857 & -0.0078\% & 44.0\% & -0.86 & 23.2 \\
& 6-h & 286 & -0.0254\% & 41.8\% & -0.38 & 78.9\\
& 12-h & 143 & -0.0217\% & 41.5\% & -0.14 & 5.6 \\
& 1-d & 73 & -0.0552\% & 44.9\% & -0.36 & 4.9 \\
\bottomrule
\end{tabular}
    \label{tab:summarystat}
\end{table}

In Table~\ref{tab:summarystat}, we report the statistical properties of BTC and ETH return series calculated from TWAP at 1-min, 5-min, 10-min, 30-min, 1-hr, 2-hr, 6-hr, 12-hr and daily sampling frequencies. 

As the interval increases, we observe that scale of mean increases naturally. At daily interval, mean return of indices is at the scale of basis points\footnote[2]{1 basis point $=$ 0.01\%}, -0.0036\% for BTC and -0.0552\% for ETH. This supports the zero conditional expectation assumption over short forecast horizon. We also observe that with time-weighted average price, the annualised volatility peaked at sampling interval of 1-hour, at $41.4\%$ for BTC and $45.3\%$ for ETH. ETH return series shows consistently higher volatility than BTC return series.

In addition, we can see negative skewness in return series for shorter sampling intervals, suggesting that negative returns are more often than positive returns when focusing on short intervals. For sampling intervals for 6-hr and 12-hr, we observed that skewness turned positive for BTC. This observation of positiveness skewness at longer sampling interval for BTC is similar to those obtained by other researchers, such as Liu and Tsyvinski~\cite{NBERw24877}. 

At all sampling intervals, we observe consistently large kurtosis for both cryptocurrencies, in excess of the kurtosis of 3 for standard Gaussian distribution. As sampling interval gets more granular, kurtosis in return series increase significantly. The evidence for leptokurtic distribution is consistent with the stylised empirical facts of cryptocurrency time series as discussed in Chi and Hao~\cite{CHI2021101421} and Tan et al~\cite{TAN2021101377}. 

\begin{scriptsize}

\begin{longtable}{llrrrrrr}
\caption{Diagnostic tests of BTC and ETH return series measured in USD. p-values are reported in brackets.} \\
%\begin{tabular}{llrrrrrr}
\toprule
\multirow{2}[3]{*}{Crypto} 
& \multirow{2}[3]{*}{Interval} 
& \multirow{2}[3]{*}{Jacque-Bera} 
& \multirow{2}[3]{*}{Dickey-Fuller} 
& \multicolumn{3}{c}{Ljung-Box}
& \multirow{2}[3]{*}{Durbin-Watson}\\
\cmidrule{5-7}
{} & {} & {} & {} & lags$=2$ & lags$=5$ & lags$=10$ & {}\\
\midrule
\endhead
\multirow{20}[3]{*}{BTC}& 1-min & 22740340.0  & -43.01  & 11380.79 & 15437.40 & 19124.01 & 1.43\\ 
  && (0.00)& (0.00)& (0.00)& (0.00) & (0.00) &\\
& 5-min & 1793332.0  & -23.55  & 1148.95 & 1692.41 & 2016.72 &2.05\\ 
  && (0.00)& (0.00)& (0.00)& (0.00) & (0.00) &\\
& 10-min & 702944.0  & -24.33  & 434.69 & 677.76 & 780.83 &1.89\\ 
  && (0.00)& (0.00)& (0.00)& (0.00) & (0.00) &\\
& 15-min & 208781.1  & -25.14 & 306.89 & 498.43 & 608.39 &1.89\\ 
  && (0.00)& (0.00)& (0.00)& (0.00) & (0.00) &\\
& 30-min & 140043.6  & -20.74 & 209.51 & 246.02 & 456.80 &2.00\\ 
  && (0.00)& (0.00)& (0.00)& (0.00) & (0.00) &\\
& 1-hr & 92473.8  & -15.24 & 8.10 & 44.63 & 115.74 &2.15\\ 
  && (0.00)& (0.00)& (0.02)& (0.00) & (0.00) &\\
& 2-hr & 14463.7 & -8.41  & 13.08 & 36.75 & 37.12 &2.10\\
  && (0.00)& (0.00)& (0.00)& (0.00) & (0.00) &\\
& 6-hr & 228.4  & -10.8  & 11.43 & 21.60 & 24.66 &2.09\\ 
  && (0.00)& (0.00)& (0.00)& (0.00) & (0.00) &\\
& 12-hr & 33.6  & -10.95  & 11.55 & 15.68 & 24.15 &1.84\\ 
  && (0.00)& (0.00)& (0.00)& (0.00) & (0.00) &\\
& 1-d & 7.00  & -8.99  & 1.99 & 5.80 & 7.85 &2.14\\ 
  && (0.03)& (0.00)& (0.08)& (0.33) & (0.64) &\\
\cmidrule{1-8}
{}& 1-min & 79131540.0  & -41.76  & 6090.82  & 7263.51  & 7827.83 &1.48 \\ 
  && (0.00)& (0.00)& (0.00)& (0.00) & (0.00) &\\
& 5-min & 2744632.0  & -24.32  & 324.49 & 477.98 & 538.76 & 1.99\\ 
  && (0.00)& (0.00)& (0.00)& (0.00) & (0.00) &\\
& 10-min & 922648.8 & -24.52  & 116.75 & 189.74 & 223.52 & 1.86\\ 
  && (0.00)& (0.00)& (0.00)& (0.00) & (0.00) &\\
& 15-min & 438395.1  & -14.88  & 88.61  & 112.52 & 132.46 & 1.88\\ 
  && (0.00)& (0.00)& (0.00)& (0.00) & (0.00) &\\
\multirow{4}{*}{ETH}& 30-min & 140427.5  & -23.92  & 65.22 & 75.32 & 169.13 & 1.97\\ 
  && (0.00)& (0.00)& (0.00)& (0.00) & (0.00) &\\
& 1-hr & 87818.2  & -28.76 & 7.41 & 21.11 & 29.30 & 2.10\\ 
  && (0.00)& (0.00)& (0.02)& (0.00) & (0.00) &\\
& 2-hr & 14640.2  & -11.68 & 5.25 & 7.45 & 7.84 & 2.01\\ 
  && (0.00)& (0.07)& (0.00)& (0.19) & (0.64) &\\
& 6-hr & 420.5  & -9.30 & 2.07 & 3.00 & 4.65 & 2.00\\ 
  && (0.00)& (0.00)& (0.35)& (0.70) & (0.71) &\\
& 12-hr & 40.9  & -8.95 & 1.45 & 3.59 & 12.70 & 1.80\\ 
  && (0.00)& (0.00)& (0.47)& (0.61) & (0.24) &\\
& 1-d & 13.1  & -9.41 & 1.87 & 3.83 & 5.79 &2.22\\
  && (0.001)& (0.00)& (0.39)& (0.57) & (0.83) &\\
\bottomrule
%\end{tabular}
    \label{tab:diagtest}
\end{longtable}
\end{scriptsize}

The non-normality in return series is further confirmed by the result of Jarque-Beta tests in Table~\ref{tab:diagtest} which reports statistically significant evidence to reject the null hypothesis of normality. These suggest the use of volatility model that captures non-Gaussian distributions.

We also applied Dickey-Fuller test to return series to test for the presence of unit root. We observe that at all sampling intervals, the null hypothesis is rejected in favour of stationarity. Moreover, serial correlations between the squared returns is tested by Durbin-Watson~(DW) statistic ranging from 0 to 4. DW $<$ 2 indicates the presence of a positive auto-correlation while DW $>$ 2 indicates the presence of a negative auto-correlation. Comparing the DW statistics to critical value at 5\% significance level, we observe statistically significant evidence for positive auto-correlation at 1-min sampling interval. For other longer sampling intervals, DW statistics range between 1.8 and 2.2 for both BTC and ETH, which does not allow us to conclude on the nature of serial correlation.

Furthermore, we applied Ljung-Box test on squared return series to assess the evidence of auto-correlation of up to 2, 5 and 10 lags respectively. We observed statistically significant evidence for serial correlation up to 12-hr interval for BTC and up to 2-hr interval for ETH . This demonstrates the existence of volatility clustering effect in the return series of BTC and ETH at finer sampling intervals, as illustrated in Figure~\ref{fig:acf}. This autocorrelated feature in squared return series motivates the inclusion of memory component in volatility forecasting when using intraday measures.

\begin{figure}[htb]
\centering
\subfloat[1-min TWAP]{%
  \includegraphics[clip,width=0.5\hsize]{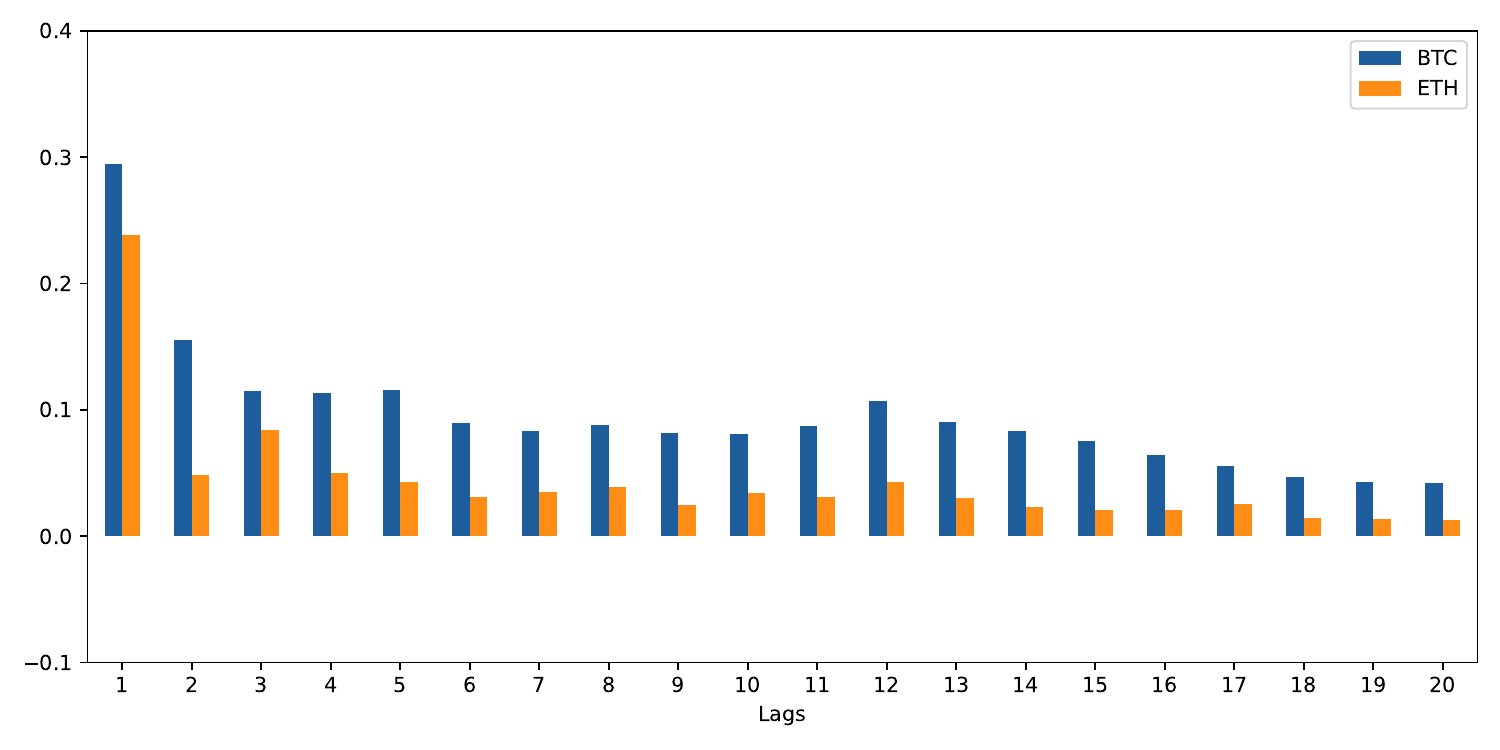}%
}
\subfloat[5-min TWAP]{%
  \includegraphics[clip,width=0.5\hsize]{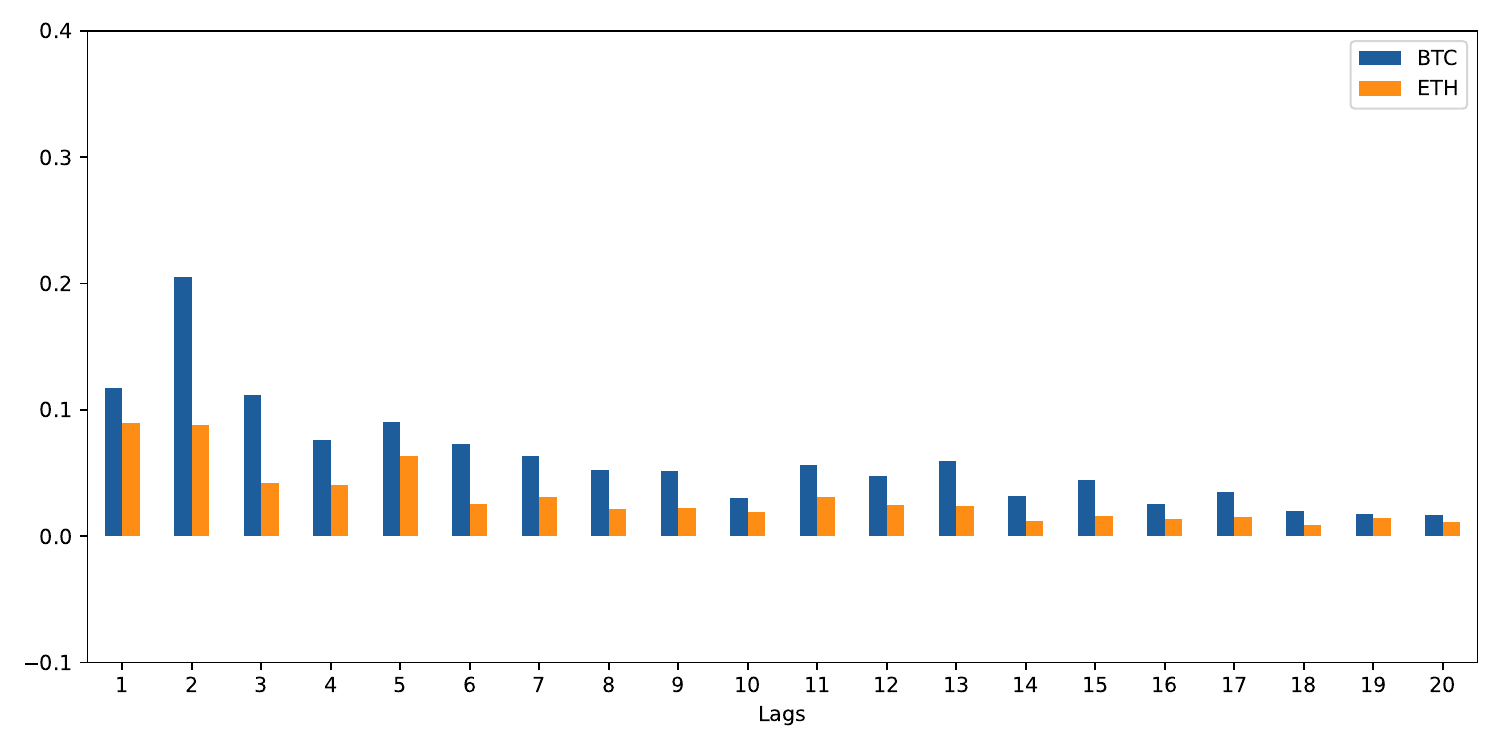}%
}

\subfloat[10-min TWAP]{%
  \includegraphics[clip,width=0.5\hsize]{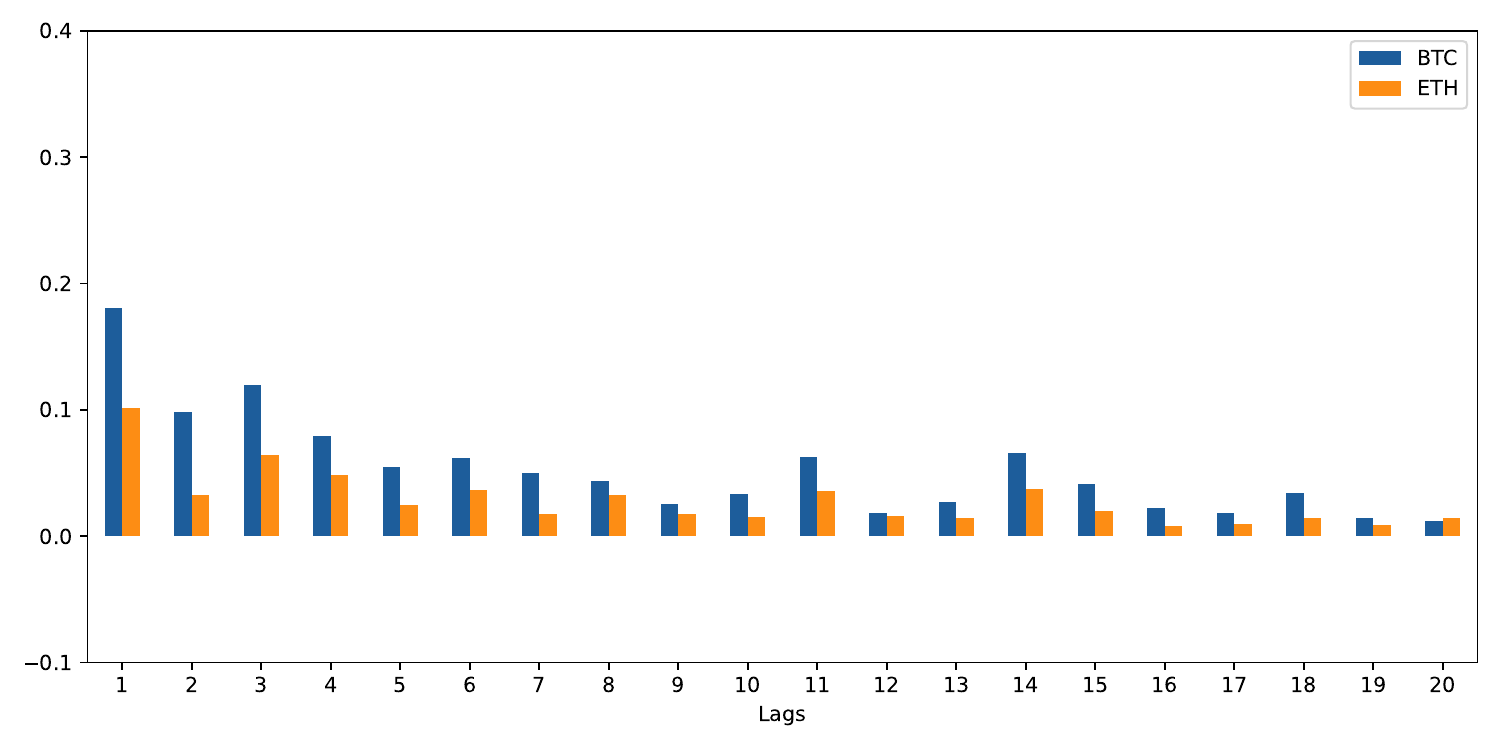}%
}
\subfloat[15-min TWAP]{%
  \includegraphics[clip,width=0.5\hsize]{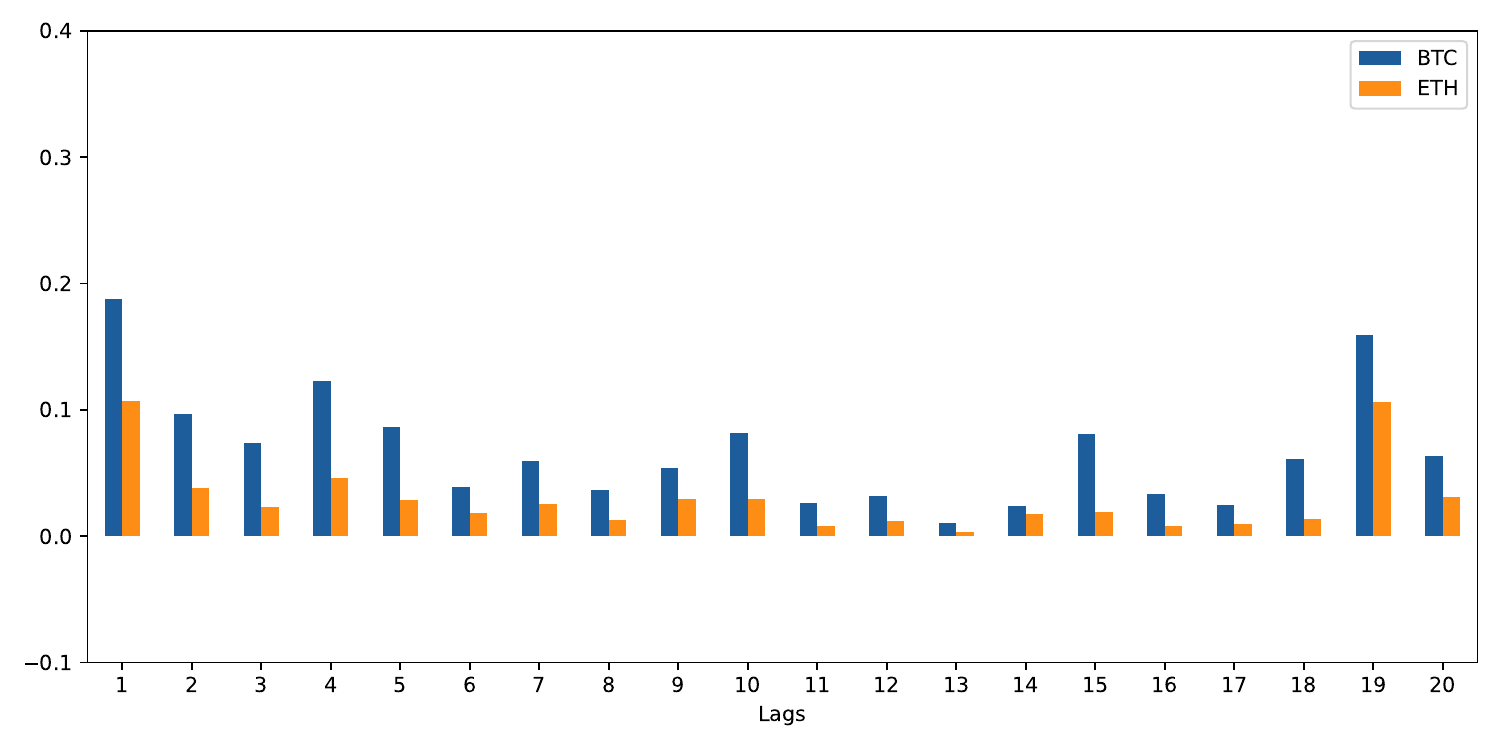}%
}

\subfloat[30-min TWAP]{%
  \includegraphics[clip,width=0.5\hsize]{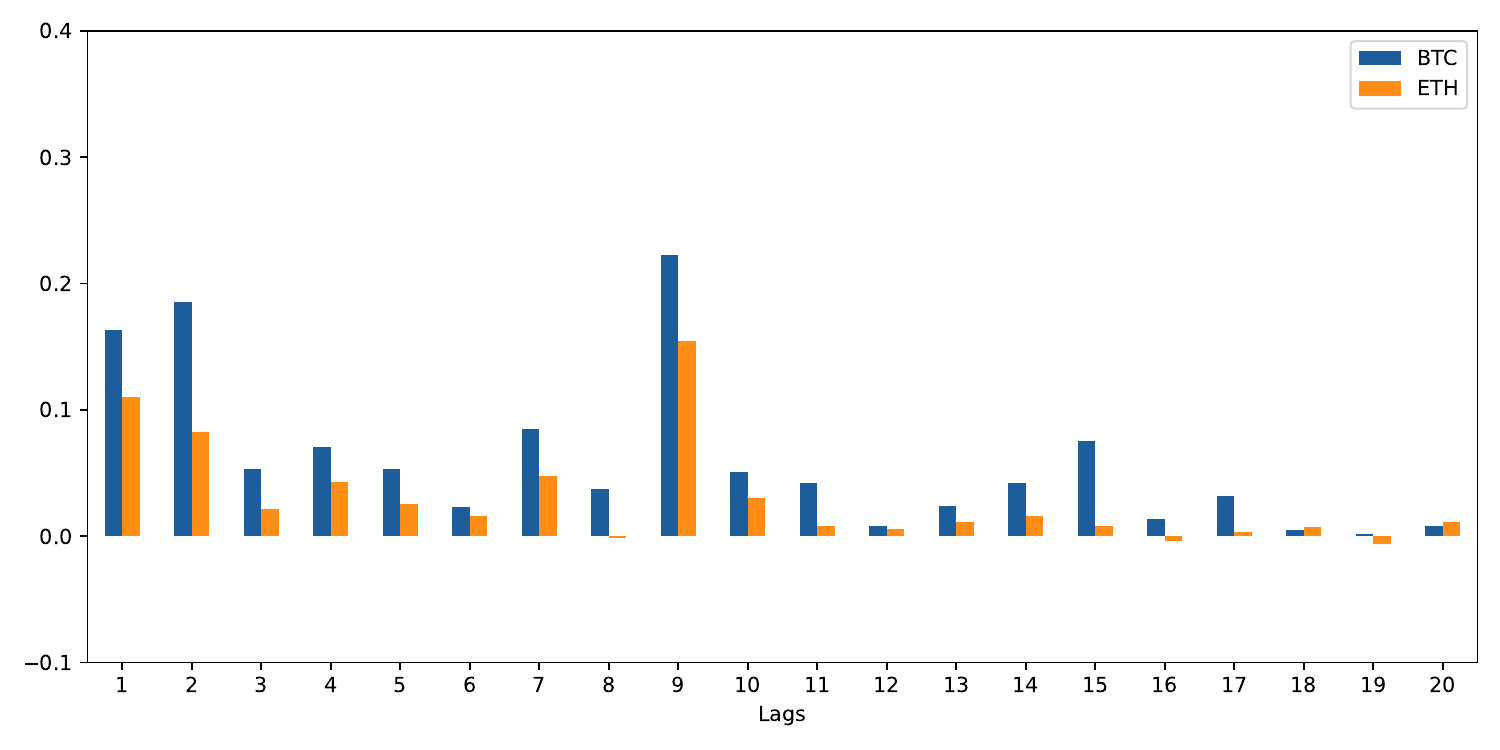}%
}
\subfloat[1-hr TWAP]{%
  \includegraphics[clip,width=0.5\hsize]{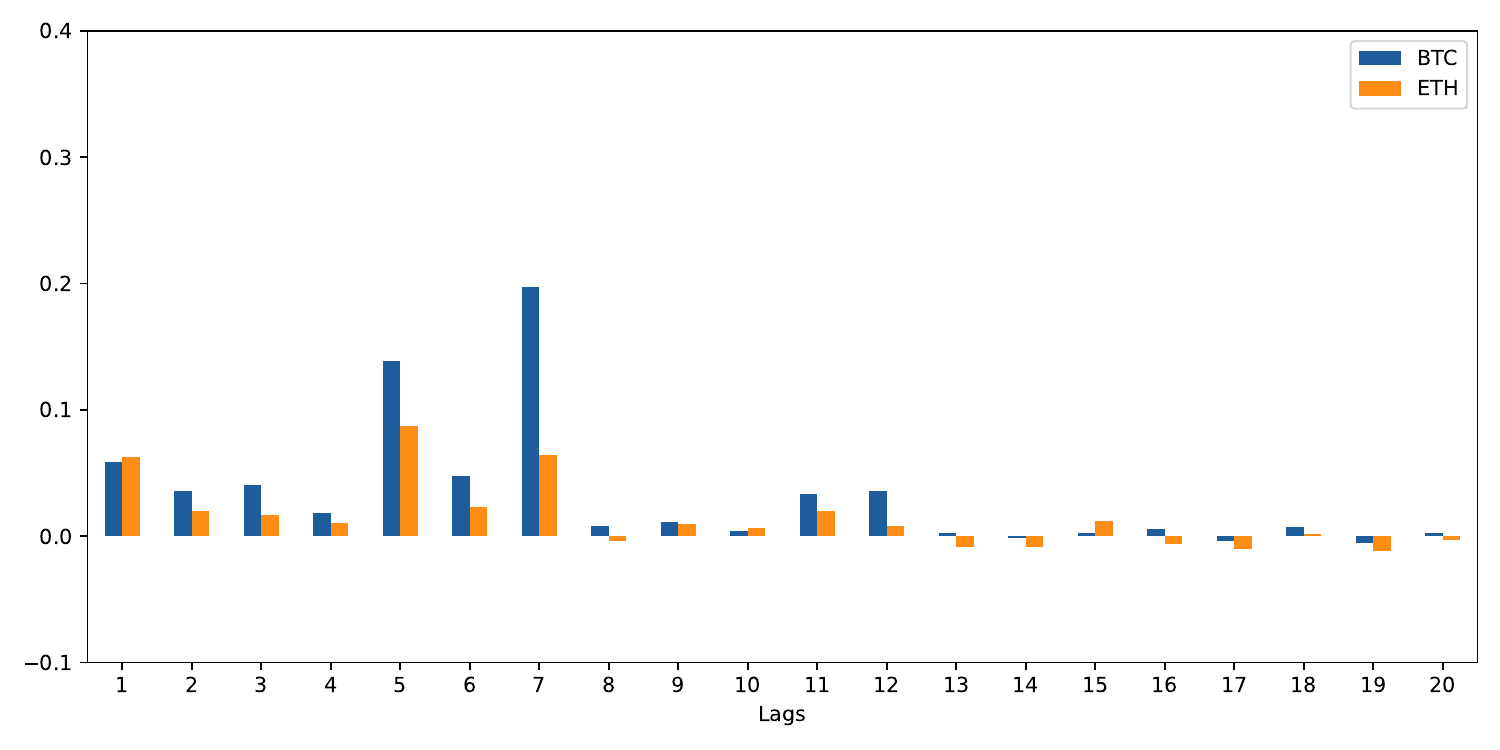}%
}

 \caption{Autocorrelations in squared returns calculated from TWAPs at different sampling intervals for BTC and ETH. Statistically insignificant $\rho$s are indicated by faded bars.}
\end{figure}

\begin{figure}[htb]\ContinuedFloat

\subfloat[2-hr TWAP]{%
  \includegraphics[clip,width=0.5\hsize]{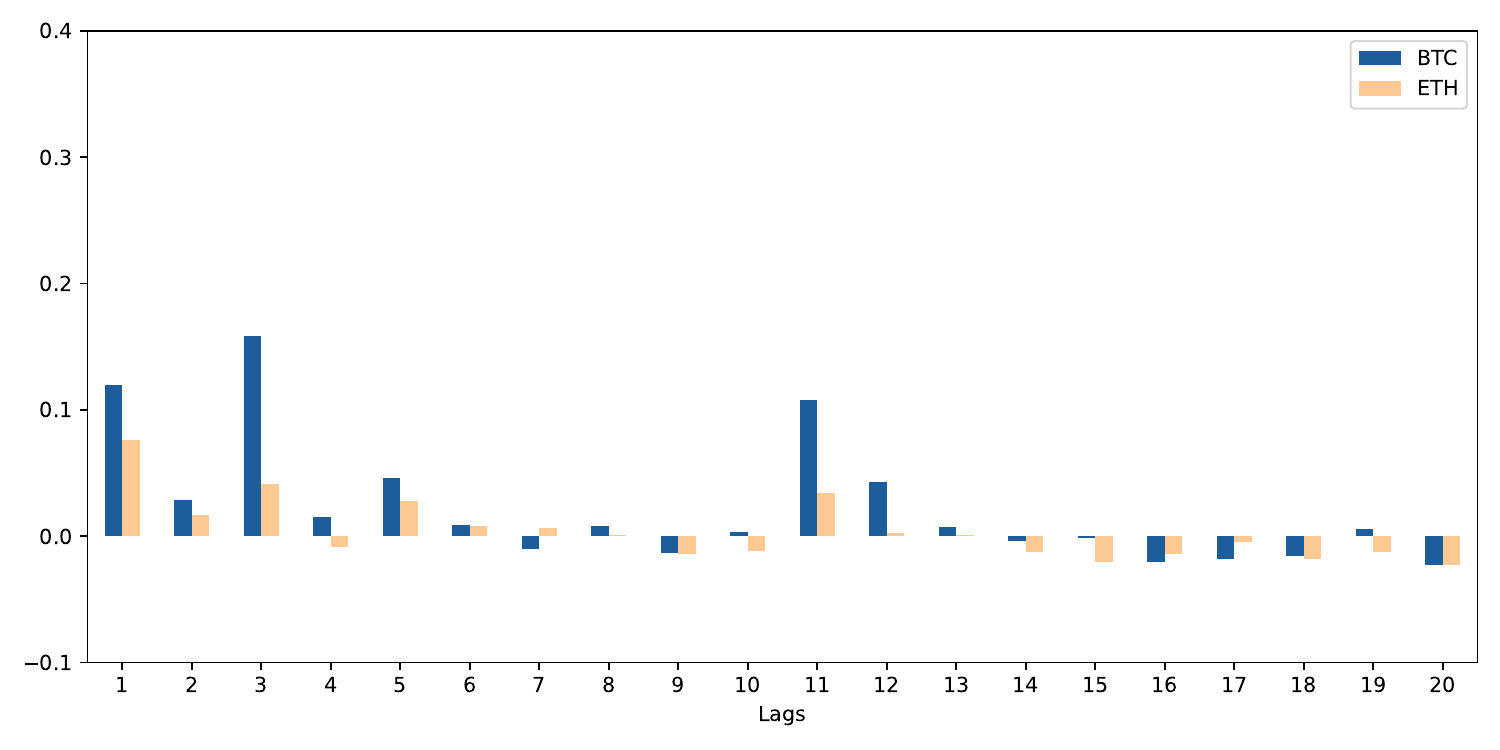}%
}
\subfloat[6-hr TWAP]{%
  \includegraphics[clip,width=0.5\hsize]{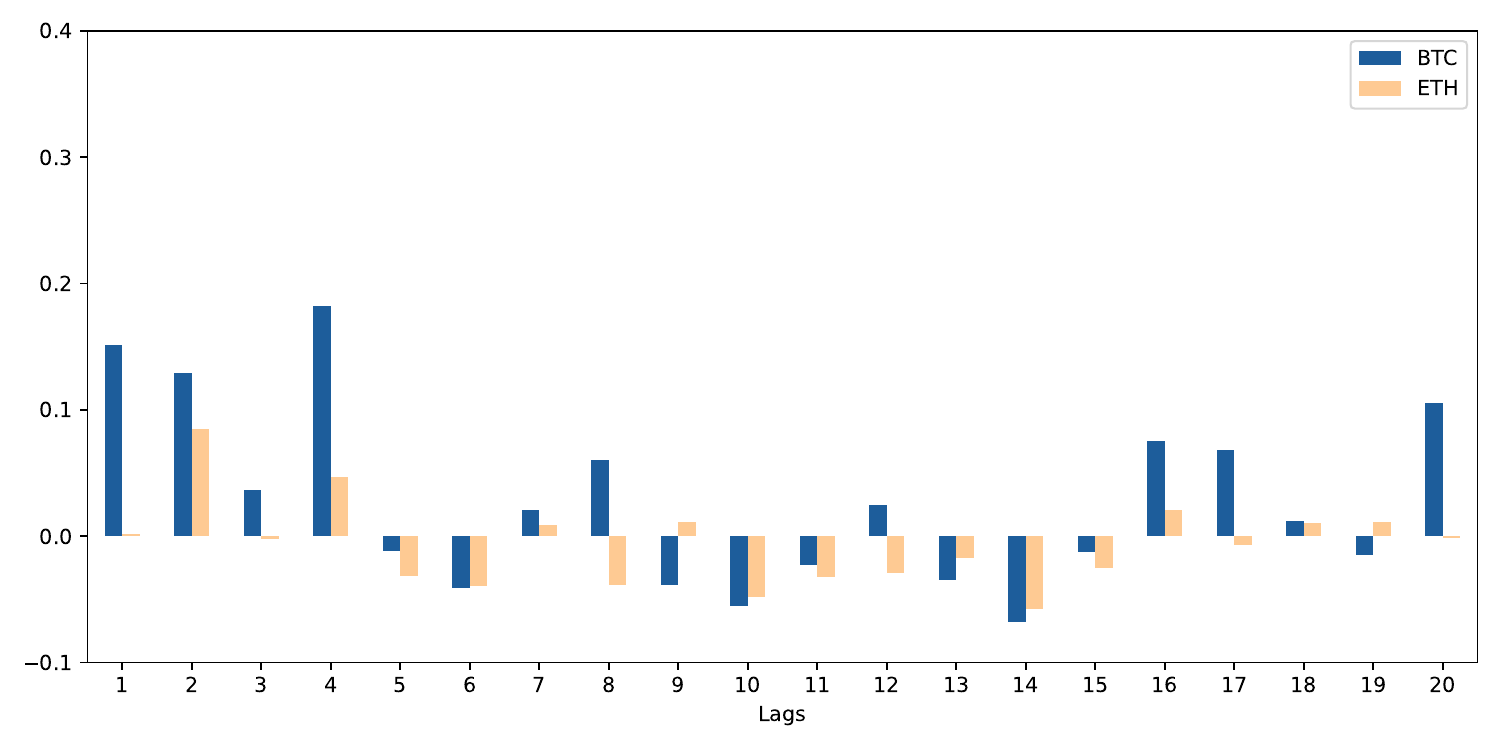}%
}

\subfloat[12-hr TWAP]{%
  \includegraphics[clip,width=0.5\hsize]{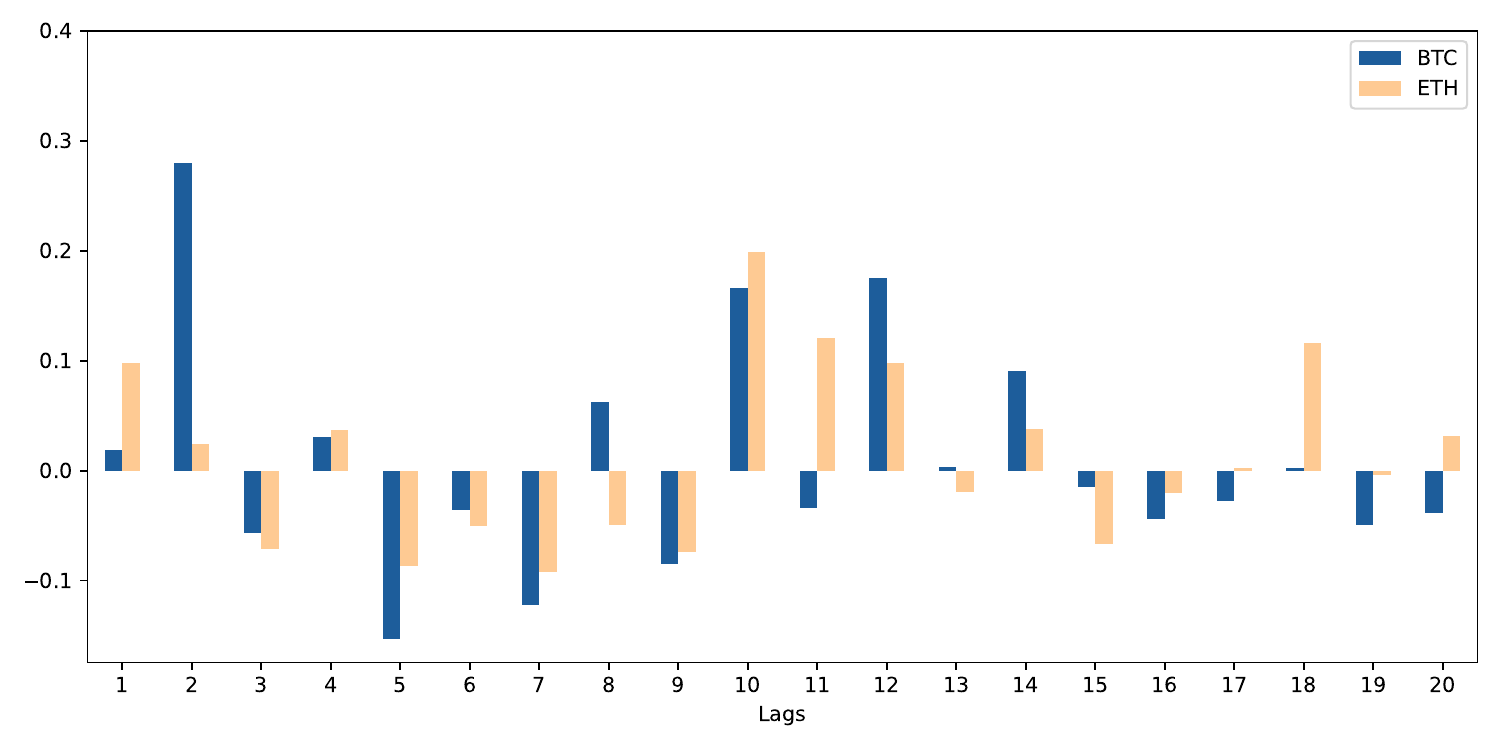}%
}
\subfloat[1-d TWAP]{%
  \includegraphics[clip,width=0.5\hsize]{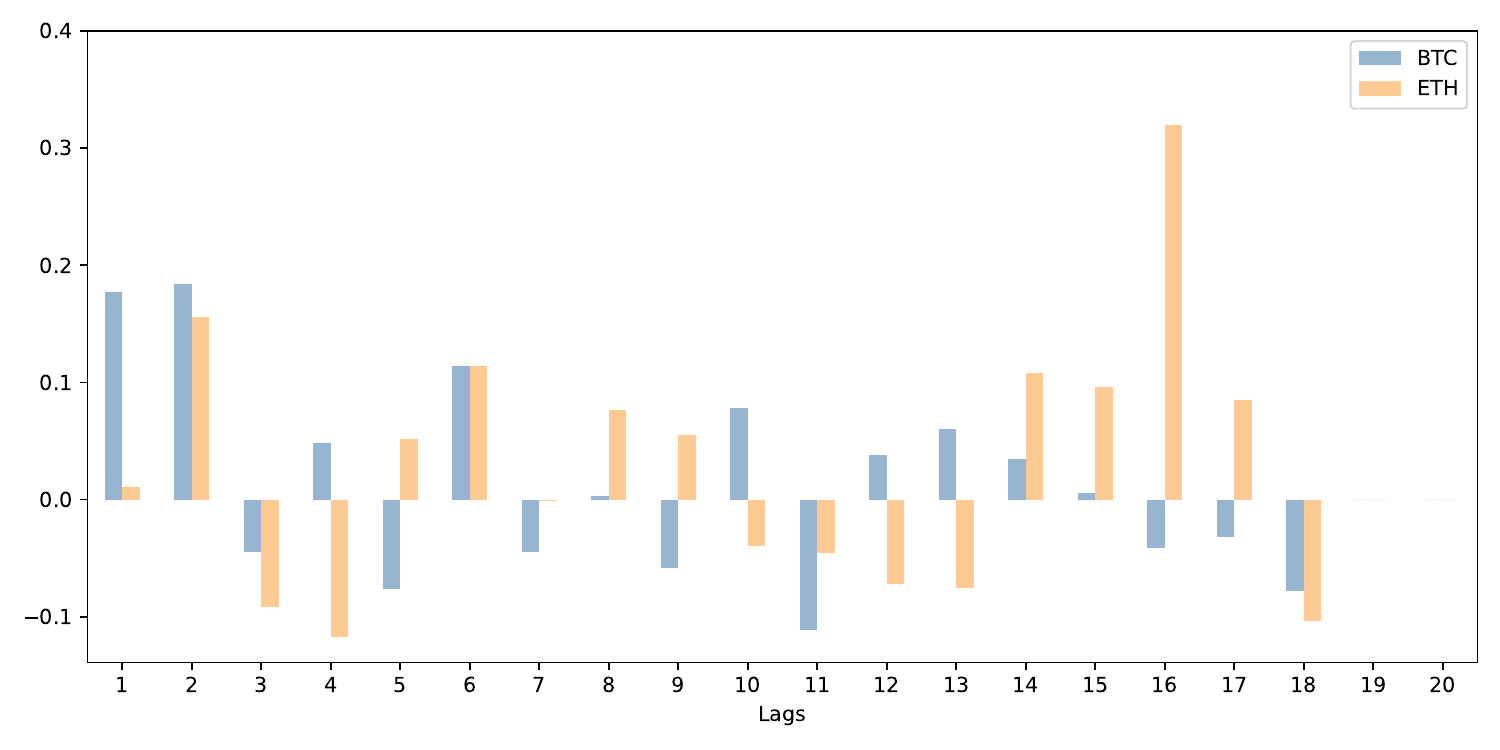}%
}
     \caption{Autocorrelation in squared returns calculated from TWAPs at different sampling intervals for BTC and ETH}
        \label{fig:acf}
\end{figure}

\begin{figure}[htb]
\centering
\subfloat[1-day realised variance]{%
  \includegraphics[clip,width=0.5\hsize]{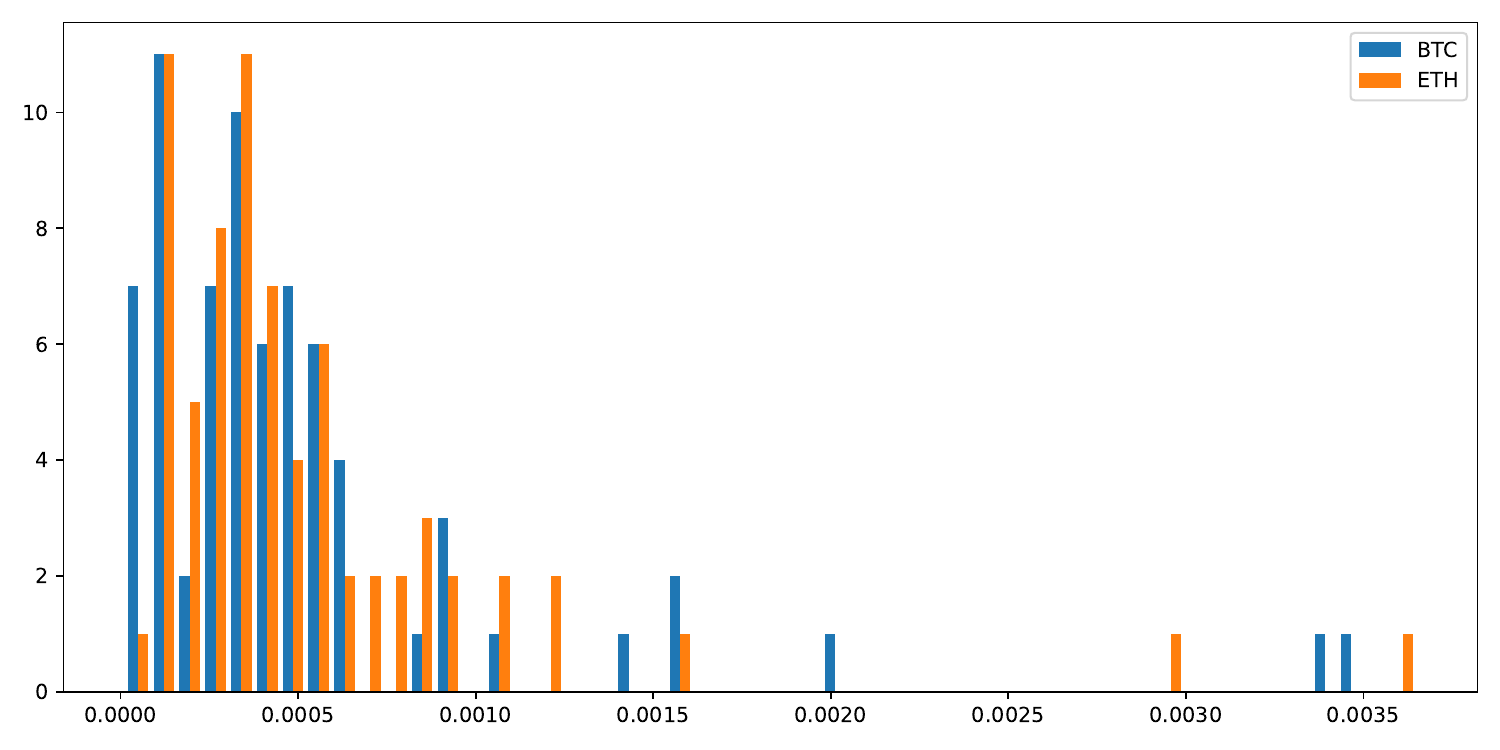}%
}
\subfloat[log-transformed 1-day realised variance]{%
  \includegraphics[clip,width=0.5\hsize]{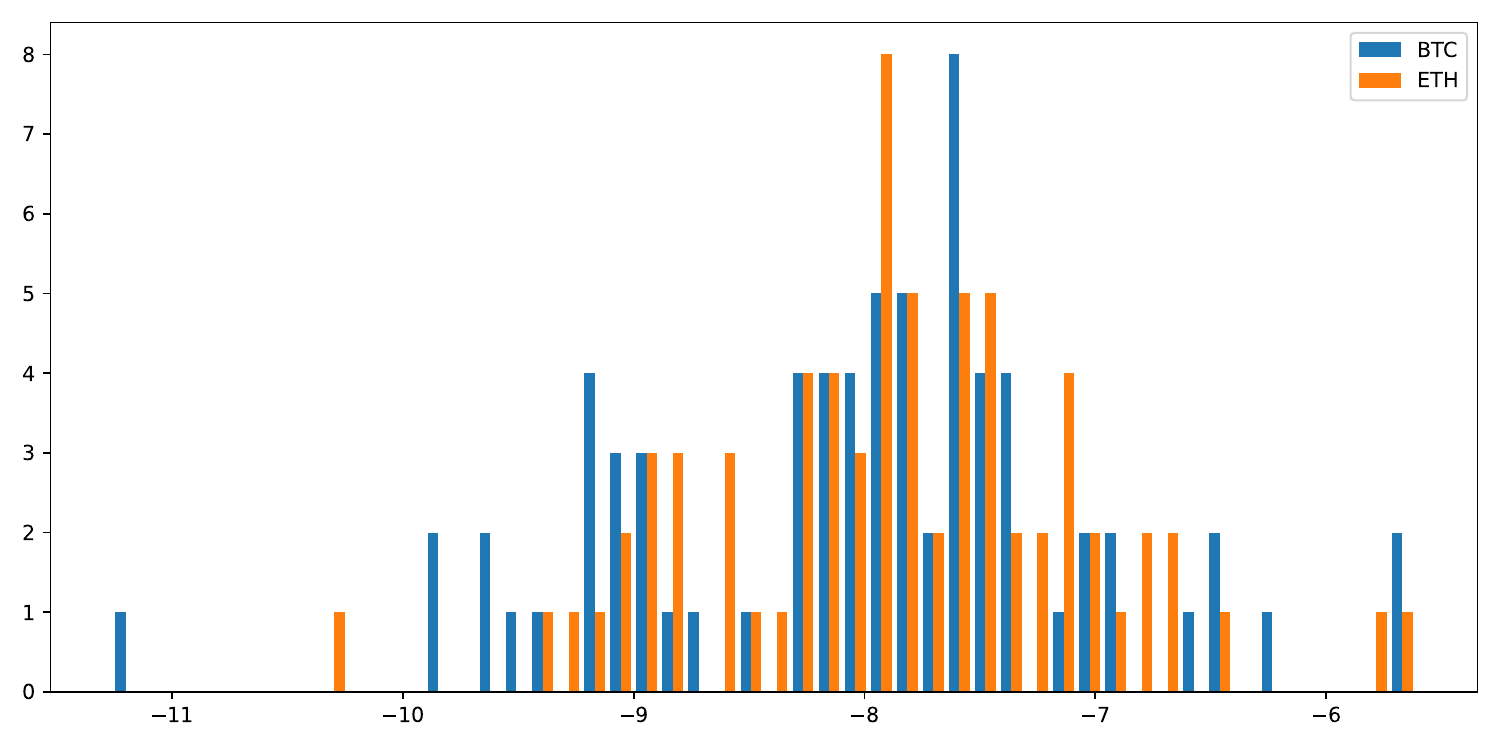}%
}

 \caption{Histograms for 1-day realised variance and log-transformed 1-day realised variance.}
 \label{fig:rvsplot}
\end{figure}

Last but not least, we analysed the statistical properties of daily realised variance and correlations of the sampled data. We observed that the realised variances for BTC and ETH significantly positively skewed, as demonstrated in Figure~\ref{fig:rvsplot}. This motivates the use of log transformation in the inference procedures later. For realised correlations, no statistically significant autocorrelation structure was observed. 

\paragraph{EWMA}
As a benchmark and also for its simplicity, we implemented the EWMA estimator to characterise the conditional distribution of return series of underlying indices. To benefit from the information in intraday data, we applied the EWMA model to return series derived from the 30-min TWAPs from previous 5 days. The choice of 30-min was to be consistent with the derivative contracts terms on Deribit, which specifies the underlying price used for settlement is the 30-min TWAP before expiry. The return series are calculated from TWAP data as natural logrithmic differences between TWAP at 30-min interval:
\begin{equation}
    r_t = \ln{\frac{{TWAP}_t}{{TWAP}_{t-30min}}}
\end{equation}

To construct the covariance matrix of underlying indices, the EWMA model of
\begin{equation}\label{eq:ewma3}\sigma_{ij, t|t-1}^2 = \lambda \sigma_{ij,t-1|t-2}^2 + (1-\lambda) r_{it}r_{jt} 
\end{equation}

are applied to different combinations of index return series. When $eps1$, $eps2$ are the same, the EWMA algorithm produces an estimation for the variance term. When they differ, the EMWA algorithm acts as an estimator for the covariance term. 

Since only BTC and ETH have tradable derivatives on Deribit and are therefore in the scope of this work, we expect the largest covariance matrix to be $2\times2$. Given the symmetricity of covariance matrix, this implies the recursions are applied to a maximum of 3 pairs of indices: btc-btc, btc-eth, eth-eth.

For the first term in the recursive function of EWMA estimator, we initialised it with the sample variance/covariance of all returns in the lookback period. This recursion detailed in Algorithm~\ref{alg:ewma} is implemented via the accumulator \texttt{over(/)} in q. 

\begin{algorithm}[htb]
    \caption{EWMA estimator}\label{alg:ewma}
    \centering
    \begin{algorithmic}
        \Require $eps1$, $eps2$, $\lambda$, $t$
        \Function{.ewma.forecast}{$eps1$, $eps2$, $\lambda$, $t$}
        \State $epscross \gets eps1*eps2$
        \If{eps1=eps2}
            \State $sigma \gets \textproc{sampleVariance}(eps1,eps2)$
        \Else 
            \State $sigma \gets \textproc{sampleCovariance}(eps1,eps2)$
        \EndIf
        \For{each item $eps$ in $epscross$}
            \State $sigma = \lambda * sigma + (1-\lambda) * eps$
        \EndFor
        \State $term \gets 48 * t * term$ \Comment{30-min estimation is scaled up to relevant forecast horizon}
        \State \Return $term$
        \EndFunction
        \end{algorithmic}
\end{algorithm}

The raw variance and covariance terms calculated from the recursion section are for a period of 30 minutes. To use it for the requested VaR calculation, we scale it up linearly to the corresponding time horizon. For example, if the VaR calculation is for $2$-day, the term will be multiplied by $96$.

In our implementation, the primary contributor to the computation duration is the iterative process involving 240 returns. If the portfolio expands to have derivatives associated with other cryptocurrency indices or digital assets, the decisive factor impacting the complexity of computations will be the quantity of underlyings. This is because the number of recursion for each term in the covariance matrix will remain fixed but the number of terms in the covariance matrix grows quadratically in correspondence with the number of underlyings. While this increase in dimension could bring challenges in terms of computation latency, kdb+ is capable of supporting parallel processings with threads and multi-processes. In this work, we could only exploit this benefit to a minimal extent, given there are only two indices within the scope of this work.

\paragraph{multivariate GARCH}
The other common family of models to capture volatility dynamic is GARCH models. In this work, we also implemented the DCC-GARCH(1,1) model with Student-t distribution to capture the dynamic of covariance matrix and accommodate the heave-tails observed in the return series. Since the model needs to be fitted with maximum likelihood estimation, we employed the \texttt{embedPy} package which calls on a Python process and performs computations in Python. The return series and relevant parameters are wrapped as Python objects which then get processed by the \texttt{mgarch} package that leverages \texttt{scipy.optimize} to minimise the negative log likelihood function of the model.

Taking into consideration that the underlying price used in settling derivative contracts is the 30-min TWAP of corresponding index before expiry, returns series used for model fitting are based on 30-min TWAP of the underlying indices. Consequently, the covariance matrix forecast will apply the same processing to TWAP data to obtain return series and at the end of estimation, goes through the same scaling process as EWMA model to adjust for the appropriate VaR time horizon.

\paragraph{HAR}\label{rv}
As an alternative to the previous inference algorithms which attempt to model the dynamics of return series, the Hetergeneous Autoregressive~(HAR) model is used to capture the volatility dynamics directly. 

For realised variance to be used as a consistent estimator of quadratic variance~(QV), an important assumption is a frictionless market, or in stochastic process term, the log-price process must be a continuous semimartingale. However, as sampling frequency increases, this assumption is violated due to the presence of market microstructure noise, such as bid-ask spreads, rounding errors, price discreteness, etc~\cite{Bandi}. To attenuate the impact of microstructure noise, we adopted the pre-averaging approach by Podolskij and Vetter~\cite{Podolskij}. In this approach, we see the log price semimartingale $P_t$ as a latent process and the observed price process $X_t$ include a noise term $\epsilon_t$:
\begin{equation}
    X_t = P_t + \epsilon_t.
\end{equation}

When we average the observed prices around $t$, the noise term of the averaged price has a lower variance than those of individually observed prices. Thus using the averaged price to calculate realised variance will produce an estimate that is closer to the optimal estimate obtained from the true latent semimartingale. In our implementation, this averaging process is delivered in the real-time price subscriber discussed in Section~\ref{subscriber} which pre-processes high frequency tick level data to the corresponding 1-min TWAP.

To estimate RV as a variance proxy, we used the return series sampled at 5-minute intervals, such that the realised variance over a period of $m$ minutes can be computed as 
\begin{equation}
    RV_{t+1}^{m} = \Sigma_{i=1}^{m/5}r_{t+i}^2, 
\end{equation}
where $r_{t+i}$ is the $i$th 5-minute return within the time interval of $m$ minutes. In the case of 12-h RV, $m = 144$.

Similarly, the realised covariance and realised correlation for underlying indices $1$ and $2$ over a period of $m$ minutes can be computed as 
\begin{equation}
    RCov_{12,t+1}^{m} = \Sigma_{i=1}^{m}r_{1,t+i}r_{2,t+i} 
\end{equation}
and 
\begin{equation}
    RCorr_{12,t+1}^{m} = \frac{RCov_{12,t+1}^{m}}{\sqrt{RV_{1,t+1}^{m}RV_{2,t+1}^{m}}}.
\end{equation}
respectively.

To forecast the covariance matrix, we adopted a modified version of HAR-DRD model proposed by Oh and Patton~\cite{OH2016349}. In the original model, the covariance matrix is decomposed into diagonal matrix of variance and correlation matrix. Each individual log-variance elements is estimated with separate univariate HAR models and the correlation matrix adopts a DCC-type estimator. The covariance matrix forecast is obtained by compositing these forecasts. While we kept the decomposition approach in our implementation, we employed HARQ model with leverage term~(LHARQ) to perform forecast for variance terms and HAR model to forecast for correlation terms as detailed below. 

For variance terms, we specify the LHARQ model using 12-hr RV as
\begin{equation}\label{lharq}
\begin{split}
    logRV_{t+1} = &\beta_0 + \beta_1r_t^{-} + \beta_2{logRV_{t}}+ \beta_3log(\sqrt{RQ_{t}}{RV_{t}}) \\ &+\beta_4\overline{logRV_{t,t-1}} + \beta_5\overline{logRV_{t,t-4}} + \epsilon_t
\end{split}
\end{equation}

,where $r_t^{-} = $min$(r_t, 0)$, $\overline{logRV_{i,j}}$ represents the log of averaged 12-hr RV from $t=j$ to $t=i$, ${RQ_{t}}$ is the realised quarticity term employed to account for estimation error in RV terms, defined as~\cite{BOLLERSLEV20161} 
\begin{equation}
RQ_{t+1}^{m} = \frac{m/5}{3}\Sigma_{i=1}^{m/5}r_{t+i}^4.    
\end{equation}

Instead of the standard HAR approach which models daily RV directly, we applied it to model the 12-hr RV, which is then scaled up to the relevant forecasting periods. This design was motivated by the analysis on the autocorrelations of squared return series shown in Figure~\ref{fig:acf}. When returns were sampled at 
12-hr interval, there was statistically significance evidence for serial-correlation; however, when we extend the sampling period to 1 day, the evidence became insignificant. 

We use the last 12-hr RV and the average of 12-hr RV over past 2 and 5 days to parsimoniously captures the high persistence in volatility. In addition, the logarithmic transformation is applied to ensure the positiveness of forecasts. At the same time, logged transformed RVs are closer to standard Gaussian distribution based on metrics of skewness and kurtosis, therefore more suitable for using OLS estimators. The inclusion of the leverage term was motivate by the empirical evidence obtained when applying the model with and without leverage term to the data used for exploratory analysis: firstly, the inclusion of leverage term improved the $R^2$ goodness-of-fit from 0.281 to 0.394 for BTC and from 0.182 to 0.245 for ETH; secondly, the coefficient of the leverage term is statistically different from 0 at 95\% confidence interval. These evidence are consistent with the aforementioned work by Yu~\cite{YU2019120707} which found that leverage effect has significant impacts on the future BTC volatility.

In order to use this model for forecasting, we fit it using OLS estimator. The parameters are estimated by solving the problem of minimisation for the sum of squared error: 
\begin{equation}
    \hat{\beta} = \underset{\beta_0, \beta_1, \beta_2, \beta_3, \beta_4, \beta_5}{\mathrm{argmin}}\Sigma(logRV_t - \widehat{logRV_t})^2
\end{equation}

A key advantage of using OLS estimator is the existence of a tractable representation of estimated parameters. In the context of implementation, as each VaR calculation is triggered, this model is fitted with OLS estimator in real-time using the built-in \texttt{lsq} function in q which leverages Cholesky decomposition for matrix inversion. 

For correlation terms, we do not apply any transformation to the realised correlation data as its distribution is not significantly skewed nor displaying heavy tails. The HAR model for realised correlation is specified as
\begin{equation}
    RCorr_{t+1} = \beta_0 + \beta_1{RCorr_{t}}+ \beta_2\overline{RCorr_{t,t-1}} + \beta_3\overline{RCorr_{t,t-4}} + \epsilon_t.
\end{equation}

While we can use OLS estimator in this model directly, there is no constraint imposed on the range of forecasted $RCorr_{t+1}$. We noted instances where the forecast failed outside the valid range of -1 to 1. As a simple remedy, we implemented an additional check in the forecast process. If invalid forecast for correlation is produced, we replaced it with the average realised correlation over past 5 days. The correlation forecast is transformed to covariance forecast by multiplying the squared roots of the corresponding variance terms.

\paragraph{Real-time inference with caching}
Recall that the primary objective of the system is to facilitate the real-time calculation of portfolio VaR. For a more accurate representation of the current market conditions, inference procedure should be executed every time a VaR calculation is triggered, incorporating the latest information from the market. In the context of cryptocurrencies, the market operates on the basis of 24/7. The conventional notion of daily opening and closing is of limited relevance within the scope of this work. When we refer to ``querying historical market data of $t$ days", we are specifically referring to data spanning the past $24t$ hours from the present moment. 

In the case of HAR model, real time inference includes queries to real-time subscriber for all 1-min TWAP data from the start of the day and to HDB for data pertaining to previous 15 days. Consequently, there should only be a subtle variation in the dataset employed for real-time inference between consecutive VaR calculation requests. To reduce the latency in data sourcing, we maintained a table named \texttt{cachedTwapHAR} in calculation process to serve as cache. Each time when an inference procedure runs, it first queries the local table for data within the previous 15 days. If there were any remaining data needed, it then queried the price subscriber and HDB only on those remaining data that did not exist in \texttt{cachedTwapHAR} yet. Considering the case where inferencing for both BTC and ETH is required, while the first inference request would take about 70ms to source all the data and save them in cache, the subsequent requests take 14ms on average.

\subsubsection{Mapping Procedure}\label{mapping}
The mapping procedure is responsible for identifying the relation between portfolio returns and underlying indices returns. Below we introduce the mapping algorithms used for a single holding, followed by an aggregated version which is implemented for the entire portfolio. The entire mapping algorithm is to produce $\tilde{\delta}$, $\tilde{\Gamma}$ and $\tilde{\theta}$ defined later in Equation~\ref{eq:portfolioreturn}. 

\paragraph{Holding Level}
For linear products, changes in holding value can be represented as a linear function of change in underlyings values. Specifically in the case of crypto futures products, this mapping has a coefficient of 1.
\begin{equation}\label{eq:futurereturn}
    V_{t+\tau} - V_t = P_{t+\tau} - P_t 
\end{equation}

For non-linear products, such as crypto options, we use a quadratic mapping whereby the change in options value is approximated via Taylor series expansion of order 2. This is also known as delta-gamma approach~\cite{Pritsker1997}. Since the value of options decreases naturally as time passes, we further included theta factor in this mapping procedure. 

Change in options value is represented as a quadratic function of changes in underlying value
\begin{equation}
    V_{t+\tau} - V_t = \delta(P_{t+\tau} - P_t) + \frac{1}{2}\Gamma(P_{t+\tau} - P_t)^2 + \theta\tau 
\end{equation}
Denoting the option return and underlying return as $r_t$ and $R_t$ respectively, the above can be expressed in returns terms~\cite{RiskMetric}
\begin{equation} \label{eq:optionreturn}
    r_t = \delta \frac{P_t}{V_t}R_t + \frac{1}{2}\Gamma\frac{P_t^2}{V_t}R_t^2 + \theta\frac{\tau}{V_t}
\end{equation}

Here we note that linear mappings in Equation~\ref{eq:futurereturn} can be expressed as a simplified version of Equation~\ref{eq:optionreturn} with $\delta =1$, $\Gamma=0$ and $\theta=0$.

\subsubsection{Portfolio Level}
For a portfolio of $n$ holdings, the portfolio return is a weighted average of returns on each holding
\begin{equation}
    r_{p,t} = \sum_{i=1}^{n}w_{i}r_{i,t}
\end{equation}
where
\begin{equation}
    w_i = \frac{V_i}{\sum_{i=1}^{n}V_i}
\end{equation}

Defining the coefficient terms in Equation~\ref{eq:optionreturn} as following~\cite{RiskMetric}
 \begin{align}
  R_t & = \begin{bmatrix}
      R_{1,t} & R_{2,t} & \cdots & R_{n,t}
  \end{bmatrix}^T \nonumber\\
  \tilde{\delta} &= \begin{bmatrix}
           w_1\frac{P_1}{V_1}\delta_{1} &  w_2\frac{P_2}{V_2}\delta_{2} & \cdots & w_n\frac{P_n}{V_n}\delta_{n}
         \end{bmatrix}^T \nonumber\\
  \tilde{\Gamma} &= \begin{bmatrix}
     w_1\frac{P_{1,t}^2}{V_{1,t}}\Gamma_1 & 0 & \cdots & 0 \\
     0 & w_2\frac{P_{2,t}^2}{V_{2,t}}\Gamma_2 & \cdots & 0 \\
     \vdots & \vdots & \ddots & \vdots \\
     0 & 0 & \cdots & w_n\frac{P_{n,t}^2}{V_{n,t}}\Gamma_n
    \end{bmatrix} \label{eqs:greeks} \\ 
    \tilde{\theta} &= \begin{bmatrix}
           \frac{w_1}{V_{1,t}}\theta_1 &  \frac{w_2}{V_{2,t}}\theta_2 & \cdots & \frac{w_n}{V_{n,t}}\theta_n
         \end{bmatrix}^T \nonumber\\
    \tau &= \text{time horizon for VaR calculation} \nonumber
 \end{align}

We could represent the portfolio return in matrix algebra as:
\begin{equation} \label{eq:portfolioreturn}
    r_{p,t} = \tilde{\delta}^T R_t + \frac{1}{2}\tilde{\Gamma} R_tR_t^T + \tau\sum_{i=1}^{n}\tilde{\theta_i}
\end{equation}

As shown above, $\tilde{\Gamma}$ is a diagonal matrix. Since derivative products in the scope of this work are based on a single underlying, cross gamma terms always equal to zero. Taking into account the required operation on $\tilde{\Gamma}$ in the transformation algorithm later, instead of maintaining a matrix, we used a vector to represent the diagonal elements without losing any information. 

\subsubsection{Real-time mapping with latest market data}
From the implementation perspective, the price sensitivities used in mapping algorithms are common metrics for options. They are part of the data feed from Deribit API. While these metrics are available from the price subscriber, to harness the benefit that q, as an integrated programming language, can operate on data directly and to avoid the latency brought by transferring the data between processes, especially in the case of portfolios with large number of holdings, we created two dedicated keyed tables -\texttt{LatestIndex} and \texttt{LatestProduct} - in the calculation process, to maintain the latest prices and greeks for each product and index. 

By setting it up as another subscriber to the tickerplant, the calculation process listens to updates from the tickerplant and updates the corresponding table with the latest market data. To illustrate the efficiency gain from maintaining these two tables, we compared the execution time of a simple \texttt{select} query for 1,000 products on the local \texttt{LatestProduct} table to the execution time of a remote query for the same group of products to the price subscriber process. Taking the average time of 100 executions, the local query takes merely 1ms to return while the remote query needs over 40ms to complete. Given the derivative universe on Deribit contains about 1,300 products and only prices and greeks data are needed for mapping, maintaining these two tables in the calculation process does not interfere with the calculation latency.

Following the definition in Equations~\ref{eqs:greeks}, the relevant market data get transformed to the target output $\tilde{\delta}$, $\tilde{\Gamma}$ and $\tilde{\theta}$ via the dedicated function \texttt{.VaR.adjustgreeks}. Since these coefficients will go through matrix algebra in transformation algorithm later, it would bring further efficiency gain if we could reduce their dimensions. Leveraging the fact that some derivatives in portfolio holdings may share the same underlying crypto index, we included an additional step to simultaneously compress all coefficients by aggregating entries in coefficients if the corresponding assets share the same underlying. Since we only have BTC and ETH indices within the scope of this work, all the coefficient terms can be reduced to vectors of dimension 2*1.

\subsubsection{Transformation Procedure}
One of the key objectives of this work is to minimise the latency in VaR calculation. The calculation should be completed within the time period of milliseconds. To achieve this, we require an analytical procedure to transform the characterisation of conditional distribution of market factors to that of portfolio value at $t+1$. 

Due to the presence of options positions in the portfolio, the mapping from underlying return to portfolio return is non-linear. It is difficult to characterise the distribution of $r_{p,t}$ with a tractable form. However, in order to calculate VaR of the portfolio, we do not require the probability density function or cumulative density function of the true distribution. Instead, the focus is on the specific quantiles of the true distribution, such as 1\% or 5\%. 

 To calculate the quantile function evaluated at these values, we implemented the Cornish-Fisher expansion, which estimates standardised quantile of the true distribution as a polynomial of the corresponding quantile of the standardised Gaussian distribution, with coefficients being functions of the moments of the true distribution~\cite{1685753b-262f-36e5-9f45-9df073254869}.

 The standardised $\alpha$th quantile of the true distribution can be estimated as
\begin{equation}
    z_{v,\alpha} = z_{\alpha} + \frac{1}{6}(z_{\alpha}^2-1)S + \frac{1}{24}(z_{\alpha}^3 - 3z_{\alpha})(K-3)-\frac{1}{36}(2z_{\alpha}^3-5z_{\alpha})S^2
\end{equation}
where $z_{\alpha}$ denotes the standardised Gaussian quantile, S and K denotes skewness and kurtosis respectively\footnote[2]{ For Gaussian distribution, S=0 and K=3, thus the standardised Cornish-Fisher quantile equals to standard Gaussian quantile.}. 
S and K used in the estimation can be obtained from central moments calculated for the true distribution. Since we have expressed the true distribution as a quadratic function of random variables $R_t$ in Equation~\ref{eq:portfolioreturn}, the central moments and the S and K parameters used for calculating portfolio return are as following~\cite{CASTELLACCI2003529,Amédée-ManesmeCharles-Olivier2019Cotc}:
\begin{align*} 
&\mu_1 = E[{r_{p,t}}]= \frac{1}{2}tr(\tilde{\Gamma}\Sigma_t) + \tau\sum_{i=1}^{n}\tilde{\theta_i}\\ 
&\mu_2 = E[{(r_{p,t} - \mu_1)^2}]= \tilde{\delta}^T\Sigma_t\tilde{\delta} + \frac{1}{2}tr(\tilde{\Gamma}\Sigma_t)^2 \\
&\mu_3 = E[{(r_{p,t} - \mu_1)^3}]= 3\tilde{\delta}^T\Sigma_t\tilde{\Gamma}\Sigma_t\tilde{\delta} + tr(\tilde{\Gamma}\Sigma_t)^3 \\
&\mu_4 = E[{(r_{p,t} - \mu_1)^4}]= 12\tilde{\delta}^T\Sigma_t(\tilde{\Gamma}\Sigma_t)^2\tilde{\delta} +3 tr(\tilde{\Gamma}\Sigma_t)^4 + 3\mu_2^2 \\
&S = \frac{\mu_3}{\mu_2^{1.5}} \\
&K = \frac{\mu_4}{\mu_2^2}
\end{align*}

With the compression step in previous mapping algorithm, the calculation for central moments at most involves matrix multiplication for dimension up to 2, which only brings negligible calculation latency.

The corresponding $\alpha$th quantile is then calculated from~\cite{Amédée-ManesmeCharles-Olivier2019Cotc}
\begin{equation}
    q_{v,\alpha} = \mu_{v} + \sigma_{v}z_{v,\alpha}
\end{equation}
, where $\sigma_{v}$ is the square root of $\mu_2$.
The VaR of interest is obtained by transforming return to market value as $q_{v,\alpha}V_{p,t}$.

\subsection{Visualisation}
The VaR calculation is delivered to users through published workspaces in KX dashboards, which is an interactive data visualisation tool developed by KX. It offers a seamless integration with the kdb+ processes by supporting kdb+/q queries as well as real-time streaming queries. 

For this work, we have built a workspace comprising of three tabs: \texttt{Futures} and \texttt{Options} for  displaying streamed analytics from data service component and \texttt{VaR Calculations} which serves as the user interface for the VaR calculation process. Below we introduce their functionalities in details.

\subsubsection{Streaming analytics}
\begin{figure}[htb]
    \centering
    \includegraphics[scale=0.3]{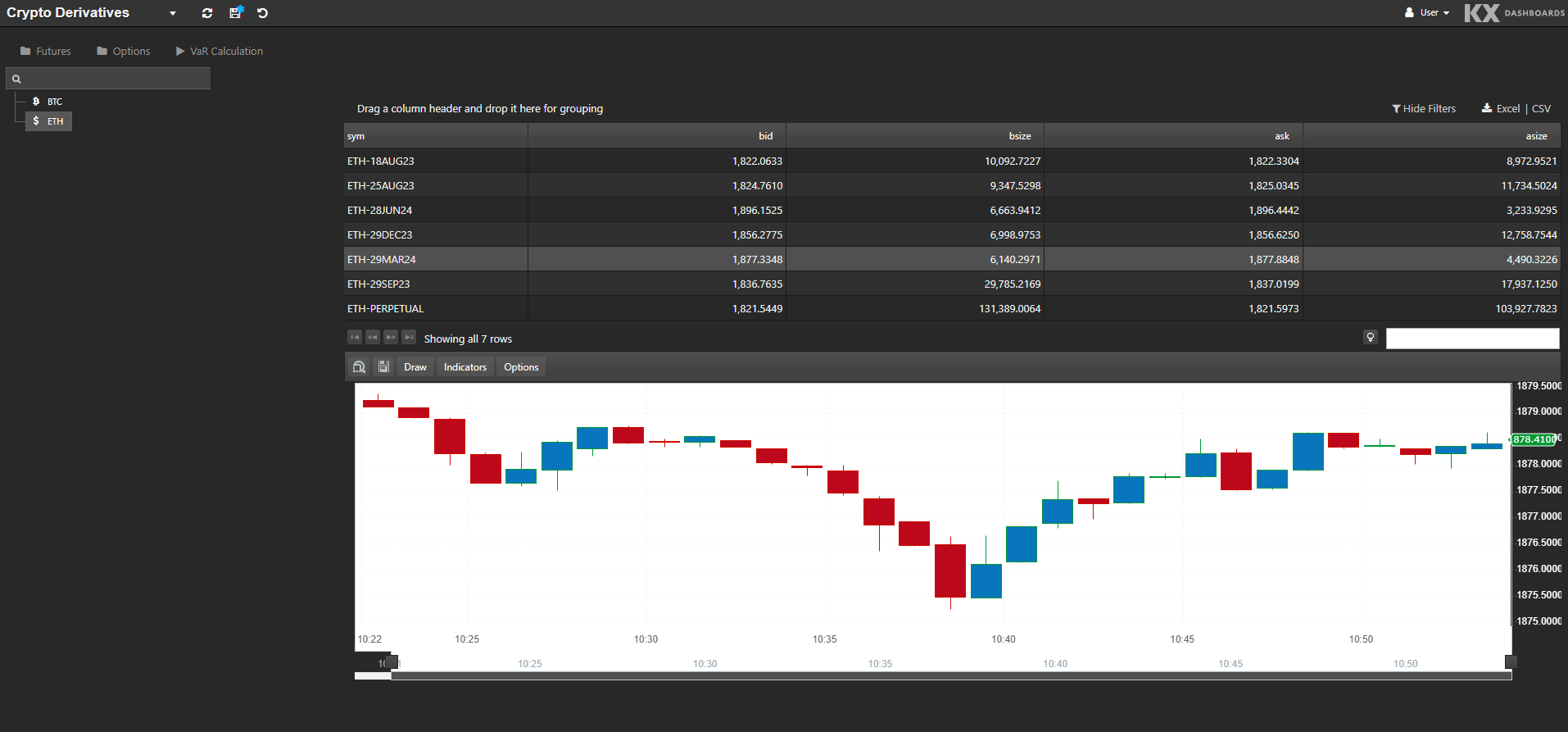}
    \caption{Workspace tab for futures}
    \label{fig:dashboardfut}
\end{figure}

\begin{figure}[htb]
    \centering
    \includegraphics[scale=0.3]{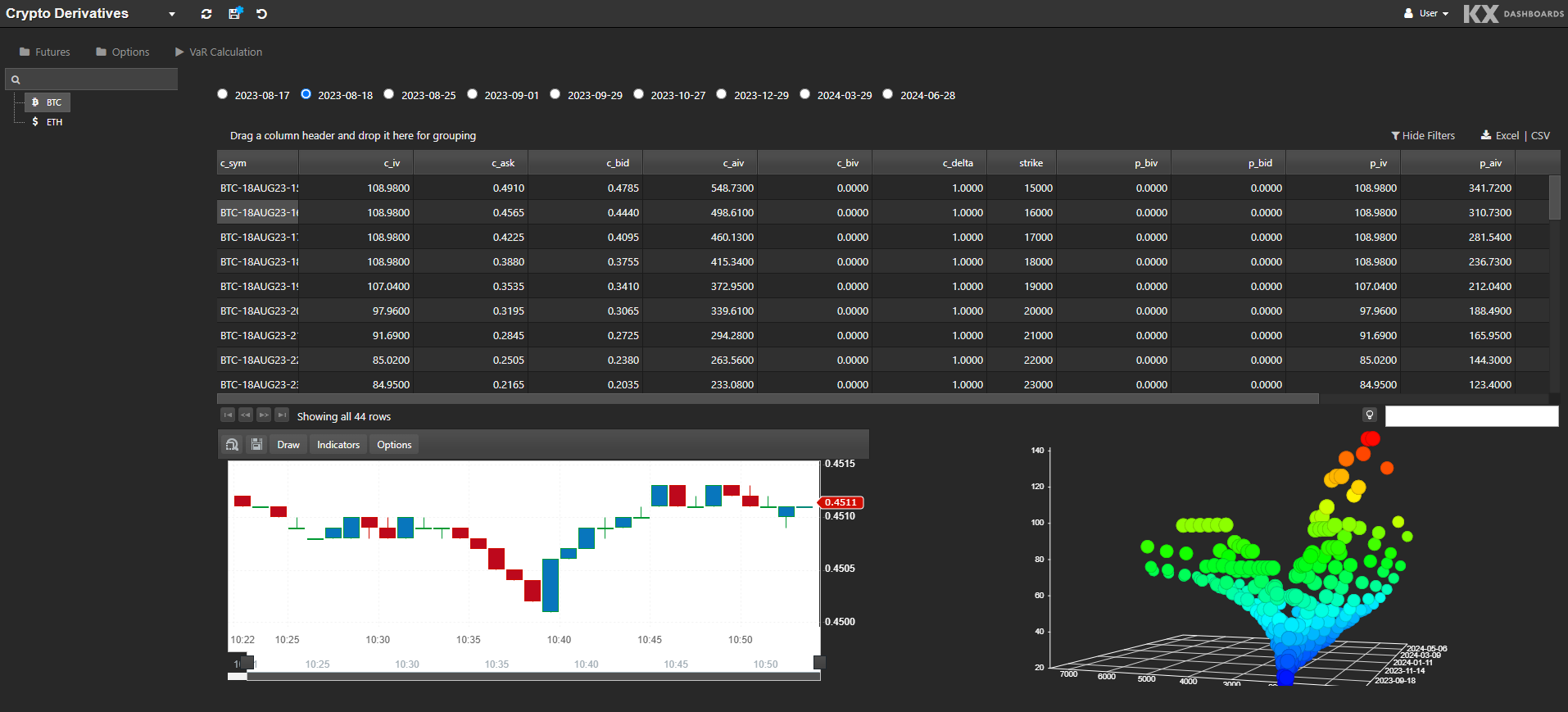}
    \caption{Workspace tab for options}
    \label{fig:dashboardopt}
\end{figure}

In order to offer contextual understanding for VaR estimates, we have created components displaying common market metrics. On a per-product basis, the tab streams information such as minute-by-minute open, low, high, close prices~(OLHC), implied volatility surface, etc. As the user selected different products from the data table as shown in the upper section of Figures~\ref{fig:dashboardfut} and~\ref{fig:dashboardopt}, the OLHC chart and the 3D volatility surface chart will update the displayed data for the corresponding products. 

The interactivity of the dashboard is facilitated through \texttt{View State} variables which store values and make them accessible across tabs and components of the workspace. Each entry in the data table is associated with a specific product ticker. Upon clicking a particular row, the product ticker is stored to the view state variable \texttt{Futures/sym} for futures tab and \texttt{Options/sym} for options tab. In the data sourcing query for OLHC chart, we implemented a dynamic query with the required view state variable as parameters. As the view state variable updates on user click, the kdb+ query update to obtain data for the corresponding product.

In both the futures and options tabs, products are initially grouped by their underlying assets, as depicted in the navigation bar situated on the left part of Figures~\ref{fig:dashboardfut} and~\ref{fig:dashboardopt}. Recognising the diverse spectrum of products available for options, an additional layer of grouping by \texttt{maturity} has been introduced. Users need to select a maturity date to view the options available at different strike levels for that maturity. Similar to the aforementioned linkage established between the data table and the OLHC chart, the radio buttons responsible for selecting maturity dates and the data table are connected through the view state variable \texttt{Options/maturity}.

\subsubsection{VaR calculation}
\texttt{VaR calculation} tab serves as the interface for the core functionality of the system, which is the real-time computation of portfolio VaR.

The workflow starts with users adding positions for their portfolios. The portfolios are identified by portfolio ID. As positions are added, the data table in the bottom half of Figure~\ref{fig:dashboardpos} updates to display the latest holdings. This automatic update is achieved through polling whereby the dashboard triggers a client-side poll of the database at a predefined interval.

\begin{figure}[htb]
\centering
\subfloat[Tab for building portfolio]{%\label{fig:dashboard2}
  \includegraphics[clip,width=\textwidth]{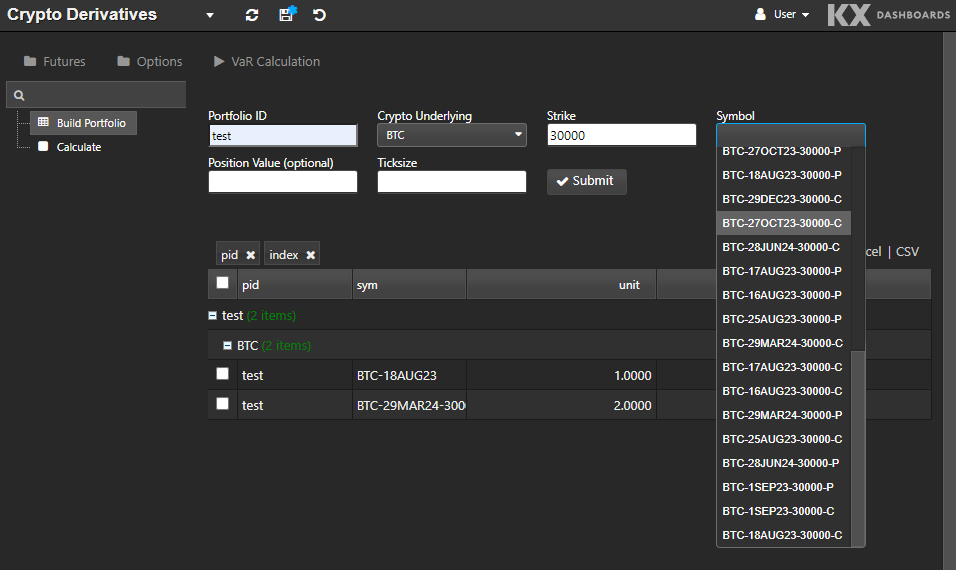}\label{fig:dashboardpos}%
}

 \caption{Workspace tab for VaR calculation}

\end{figure}

As portfolios are constructed, users switch to \texttt{Calculate} tab using the navigation bar on the left. Within the \texttt{Calculate} tab, as displayed in Figure~\ref{fig:dashboardvar}, users can select the portfolio for which the VaR calculation is required from the dropdown, then provide the confidence level and time horizon parameters together with the chosen volatility model.

Upon clicking the \texttt{Calculate} button, the dashboard triggers a request to the calculation process, calling the \texttt{.VaR.estimate} function with 4 parameters: portfolio id, confidence interval, time horizon and inference model. The choice of inference model is defaulted to \texttt{HAR}. The process will return the result as a \texttt{dictionary} atom, which is mapped to corresponding view states in the dashboard to be displayed to users.

\begin{figure}[htb]\ContinuedFloat

\subfloat[Tab for VaR Calculation]{%
  \includegraphics[clip,width=\textwidth]{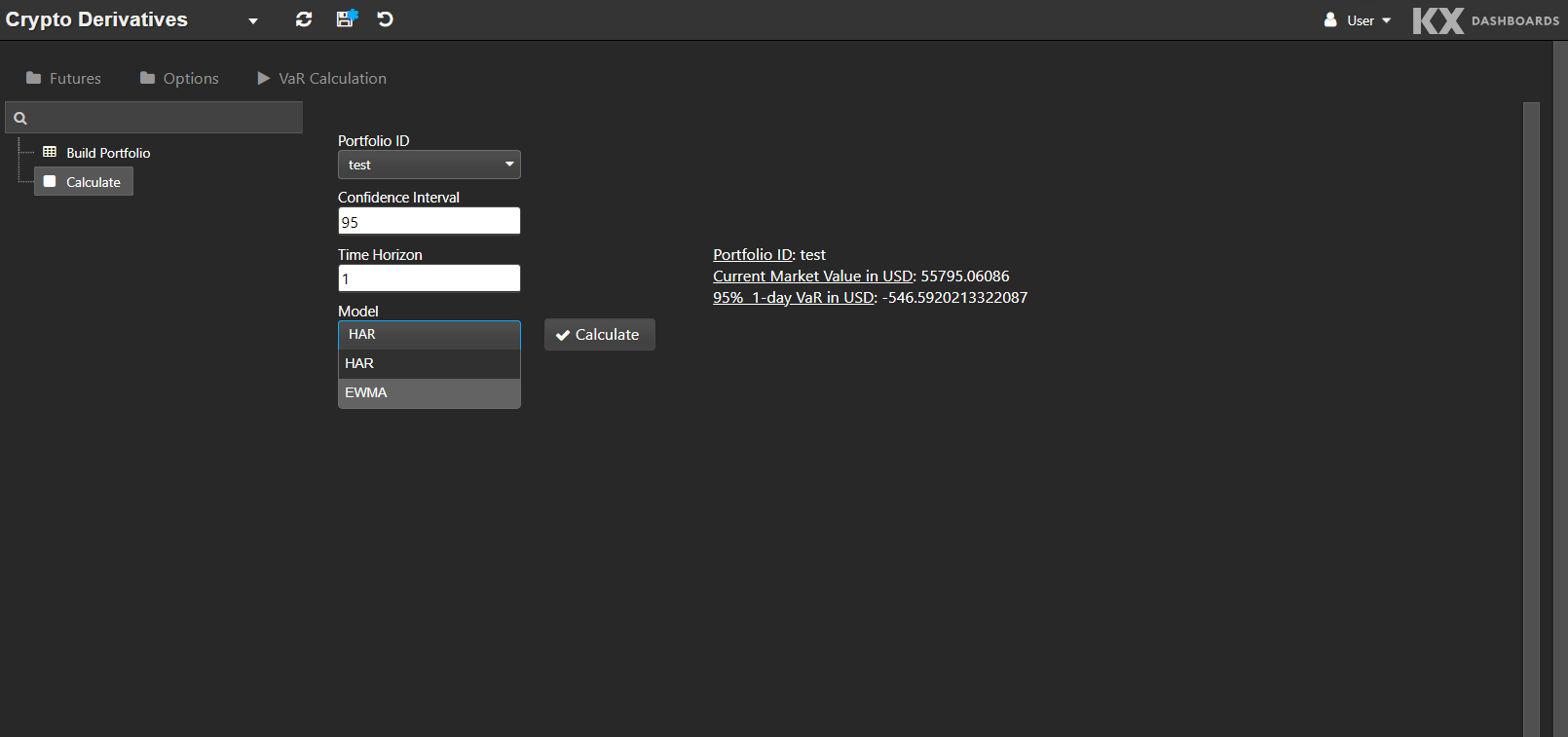}\label{fig:dashboardvar}%
}
     \caption{Workspace tab for VaR calculation}
       \label{fig:dashboard2}
\end{figure}

%%%%%%%%%%%%%%%%%%%%%%%%%%%%%%%%%%%%
\section{Evaluation}\label{chap:Evaluation}
Within this section, we will evaluate the system against the objectives of this work. The assessment has two parts: firstly, we assess the performance of the system in terms of computation latency; following that, we empirically examine the accuracy of VaR estimates using different backtest strategies. 

\subsection{Latency Performance}\label{ipc}

\subsubsection{Evaluation Setup}
For the evaluation on calculation latency, we utilised the system command \texttt{\symbol{92}ts} in kdb+ which executes the calculation and records the execution time in milliseconds and space used in bytes. The assessment encompassed portfolios of different numbers of holdings, ranging from 1 to 1000. For this evaluation, we run all the processes on a single server with AMD EPYC processor of 4 cores, 8GB memory.

As outlined in Figure~\ref{fig:proc} and Algorithm~\ref{alg:var}, the overall workflow of VaR computation includes three key calculation steps: volatility inference on underlying crypto indices, portfolio mapping through delta-gamma-theta approximation and transformation via Cornish-Fisher expansion. For this evaluation, we define the time consumption for inference to be $t_1$, the time for portfolio mapping to be $t_2$, the time for transformation to be $t_3$, and the time for other miscellaneous steps as $t_{\epsilon}$, thus we have the overall system response latency as:
\begin{equation}
    t = t_1 + t_2 + t_3 + t_{\epsilon}.
\end{equation}

\subsubsection{Result}
We started with the evaluation on the calculation latency of different inference models. Given there are only BTC and ETH derivatives available on Deribit, the maximum number of underlying indices to perform inference on is two, leaving the effective calculation complexity at $\mathcal{O}(1)$ against the number of holdings.

\begin{table}[htb]
    \small
    \centering
    \caption{Execution time in milliseconds for volatility inferencing with different volatility models, assuming the inference needs to be performed for both BTC and ETH. Data sourcing latency for HAR assumes the use of cache table introduced in Section~\ref{inference}.}
    \begin{tabular}{cccc}
    \toprule
        {Steps} & {EWMA} & {DCC-GARCH}& {HAR} \\
    \midrule
        Data Sourcing & 12.4 & 6.3 & 20.5 \\
        Inference&1.3& 7932 &18.0\\
    \bottomrule
    \end{tabular}    
    \label{tab:latency}
\end{table}

A comparative analysis of the latency of three inference methodologies shows that EWMA has superior performance than GARCH and HAR in terms of calculation latency. From the breakdown presented in Table~\ref{tab:latency}, we noted that the higher latency in HAR model is largely driven by the inference algorithm. This is caused by the additional steps to transform log returns into realised measures for inference: while EWMA and GARCH models merely requires log return data over 5 days with 48 data points per day which can be used directly in the inference and forecast steps, HAR model needs to transform per-minute data over 15 days with 1440 data points per day before using them for inference. In addition, in comparison to EWMA model which performs iterations of simple algebra, HAR model involves solving LU decompositions for model fitting and consequently has a higher computation latency. Another point to highlight is the high latency in DCC-GARCH model, which is caused by kdb+ process calling python engines to fit the model with maximum likelihood estimation.

To assess the calculation latency from mapping procedures, we broke down it into time taken to source real time market data and time taken to apply the mapping algorithm. To further illustrate the efficiency gain from maintaining the \texttt{LatestProduct} and \texttt{LatestIndex} tables in the calculation process, we compared the latency caused by using local and remote queries. As indicated in Tables~\ref{tab:latencybyipc}, as portfolio holds more products, the latency and space used for data sourcing increases naturally as more data are required, resulting in more efficiency gain for using local tables in the calculation process.

\begin{table}[htb]
    \small
    \centering
    \caption{Execution time and space used for mapping procedure with local and remote queries. Run time and space used are computed as the average of 100 times of execution and presented in milliseconds and KB respectively. $<$ where execution time is less than 0.1 ms.}
    \begin{tabular}{ccccccc}
    \toprule 
        \multirow{3}[3]{*}{Holdings} & \multicolumn{4}{c}{Data Sourcing}& \multicolumn{2}{c}{\multirow{2}[2]{*}{Mapping}}\\
        \cmidrule{2-5}
        {}&\multicolumn{2}{c}{Local} & \multicolumn{2}{c}{Remote}&{} \\
        \cmidrule(lr){2-3} \cmidrule(lr){4-5} \cmidrule(lr){6-7}
        {}&{Time}&{Space}&{Time}&{Space}&{Time}&{Space}\\
    \midrule
        1 &$<$& 12.5&21.2 &8.8& $<$ & 2.9\\
        5 &$<$&13.2 & 22.8&9.2& $<$&3.3\\
        10 &$<$&13.2 &24.6 &9.6& $<$&4.5\\
        20 &$<$& 15.1&23.9 &10.1& $<$&6.5\\
        50 &0.1& 16.6&25.6 &30.1& $<$&10.4\\ 
        100 &0.1&16.7&30.4& 57.1&0.1&18.6\\
        200 &0.2&21.1& 31.5& 59.3&0.1& 35.0\\
        500 &0.5&37.5 &38.7 &156.9& 0.3&67.7\\
        1000 &1.0&43.4& 45.9& 1256.3& 0.5&133.3\\
    \bottomrule
    \end{tabular}    
    \label{tab:latencybyipc}
\end{table}

Recall that in Section~\ref{mapping}, we compressed to coefficients used in transformation algorithm to vectors of dimension 2*1. The calculation in this step only involves simple matrix algebra, followed by the analytical formula of Cornish-Fisher expansion. The number of holdings does not affect the calculation latency in this step. We reported the average execution time for 100 times of transformation algorithm as 0.03ms.

The examination of per-stage execution times reveals that a predominant portion of the latency observed can be attributed to the inference procedures. In the context of EWMA model, almost all the latency arise from sourcing historical data for performing the inference procedures. For HAR model, the latency is evely split between data sourcing and performing inference and forecast. 

We finally tested on the full workflow calculation latency using a portfolio holding every derivative products available on Deribit, with total number of holdings at nearly 1300. Overall, with Unix socket for IPC, using EWMA model for inference produces the smallest calculation latency of 14.2ms, followed by 38.2ms for HAR model. 

\subsection{Accuracy Performance}
To assess the performance in terms of accurancy, we conducted VaR backtests with portfolios for the periods from 2023.07.25 to 2023.08.12, with 24 equally spaced timestamps on each day, totaling to 413 samples\footnote[3]{Data in certain intervals within the periods are missing, we removed samples affected by missing data, thus total number of samples is less than 19 days * 24 samples per day = 456.}. Portfolio holdings for each sample were drawn from the derivative products available in the corresponding time horizon. The backtests for VaR focus on two properties: unconditional coverage property and independence property~\cite{f089d1fa-3082-3934-9893-9bf3432ca0a9}. We could only conclude on the correctness of conditional coverage by the VaR forecast when both properties are satisfied. 

\subsubsection{Unconditional Coverage Test}
Given VaR is a downside risk metric, the focus of the evaluation is on the underestimation of risk. Let a violation event $i$ be defined as 
\begin{equation}
    i=\begin{cases}
        1, & \text{if realised loss for portfolio $i$, time horizon $t$ $\ge$ estimated VaR}\\
        0, & \text{else}
\end{cases}
\end{equation}
The most succinct test is to compare the observed number of violations to the expected number of violations through the binomial distribution test. A key assumption in this test is that violation events are identically and independently distributed, such that the total number of violations $I$ can be seen as following a binomial distribution, with $p$ corresponds to the defined quantile $\alpha$ of the VaR estimates. The null hypothesis is as:
\begin{equation}
    H_0: I \sim \mathcal{B}(413,\,p)
\end{equation}

For this backtest, we considered 95\% 1-day VaR, $97.5\%$ 1-day VaR and  99\% 1-day VaR. This leads to $p$ of $5\%$, $2.5\%$ and $1\%$ for the binomial distribution respectively. We run the VaR calculation procedure on 413 sample portfolios. In Table~\ref{tab:binomialtest} below, we report the results of the coverage test.

\begin{table}[htb]\small
    \centering
    \caption{Coverage test for VaR calculated with EWMA, GARCH, HAR estimators, benchmarked to VaR calculated with ex-post realised covariance matrix. $I_{\text{Expected}}$ has been rounded to interger. p-values for binomial test are reported in brackets.}
    \begin{tabular}{rcccccc}
    \toprule
        {VaR Level} & {Observations}&{$I_{\text{expected}}$}&{$I_{\text{realised}}$}&{$I_{\text{EWMA}}$}&{$I_{\text{GARCH}}$}&{$I_{\text{HAR}}$}  \\
    \midrule
        \multirow{2}{*}{$95\%$} & \multirow{2}{*}{413} & \multirow{2}{*}{21} & 13& 23& 33& 17\\
        {}&{}&{}&(0.974)&(0.328)&(0.006)&(0.825)\\
        \multirow{2}{*}{$97.5\%$} & \multirow{2}{*}{413} & \multirow{2}{*}{11}& 8 & 19 & 30 & 10\\
        {}&{}&{}&(0.811)&(0.009)&(0.000)&(0.58)\\
        \multirow{2}{*}{$99\%$} & \multirow{2}{*}{413} & \multirow{2}{*}{4}  & 7 & 14 & 19 &6\\
        {}&{}&{}&(0.124)&(0.000)&(0.000)&(0.235)\\
    \bottomrule
    \end{tabular}    
    \label{tab:binomialtest}
\end{table}

In general, when number of violations is higher than the expected value indicated in column $I_{\text{expected}}$, it implies that VaR measure underestimates the risk; when actual violations are lower than the expected value, it indicates overestimation. For a VaR model to be accepted at the confidence level of $95\%$, we expect the number of violations to be within range of 2 standard deviations from the expected value. 

By assessing the p-values, we observe statistically significant evidence at $99\%$ confidence level to reject EWMA and GARCH estimators for all three VaR levels. In particular, we observe that the issue of underestimation is more profound in DCC-GARCH estimator than EWMA estimator as indicated by the larger number of violations across the three VaR levels. For HAR estimator, while it showed a robust performance for $95\%$ 1-day VaR metrics, it should be rejected at $99\%$ confidence level for $97.5\%$ and $99\%$ 1-day VaR.

 Due to the known limitation of using delta-gamma-theta approach to approximate the tail behaviour of the return distribution, we created an additional benchmark by applying the ex-post realised covariance matrix to the mapping and transformation procedures. The result of this benchmark is presented in column $I_{\text{realised}}$. By evaluating the violations data for the implemented inference procedures against this benchmark, we can assess the performance of inference estimators in isolation.

The results of the coverage test is further illustrated in Figure~\ref{fig:coveragetest}. We compare the VaR estimates against the actual loss in terms of returns. Violations from EWMA, DCC-GARCH and HAR estimators as well as ex-post realised measures are indicated as coloured vertical lines above the trend lines. Notably, EWMA and DCC-GARCH violations are concentrated around 2023.07.31 and 2023.08.02. This prompts us to apply another set of test for assess the dependence between violations in the next section.

\begin{figure}[htp]

\centering
\subfloat[$95\%$ 1-day VaR]{%
  \includegraphics[clip,width=0.95\columnwidth]{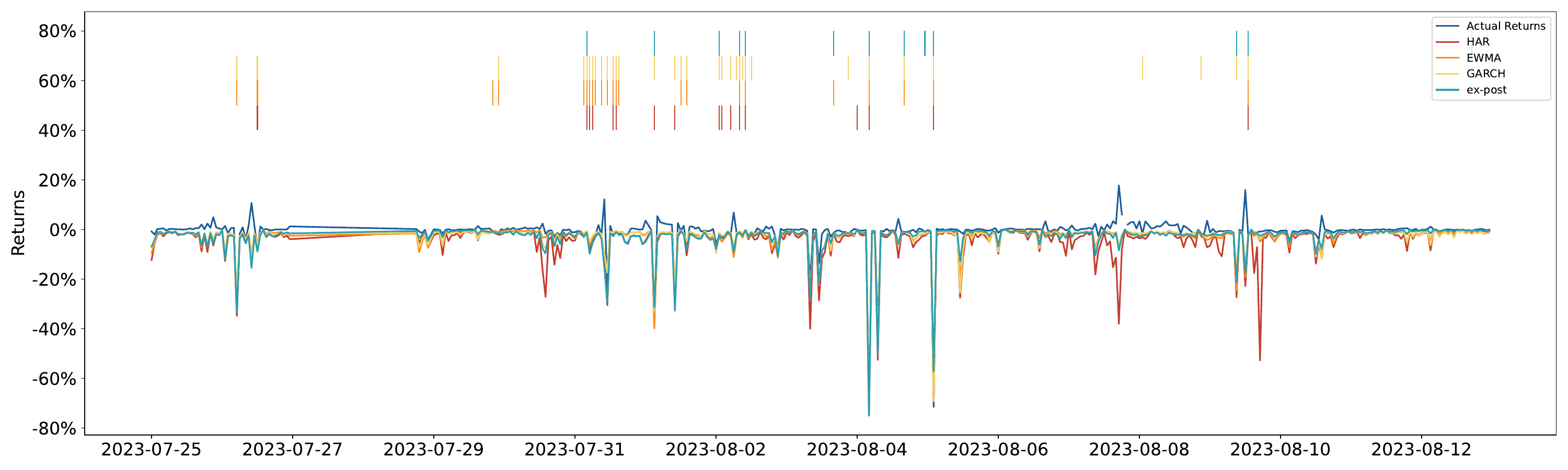}%
}

\subfloat[$97.5\%$ 1-day VaR]{%
  \includegraphics[clip,width=0.95\columnwidth]{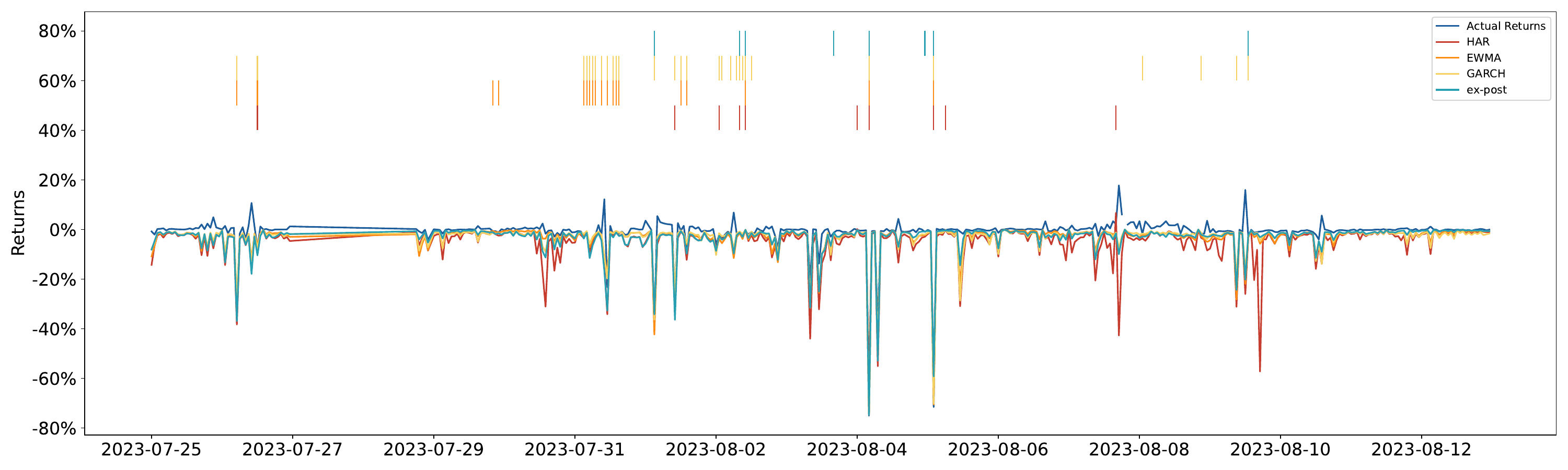}%
}

 \caption{Coverage Test for 1-day VaR}
\end{figure}

\begin{figure}[htp]\ContinuedFloat

\subfloat[$99\%$ 1-day VaR]{%
  \includegraphics[clip,width=0.95\columnwidth]{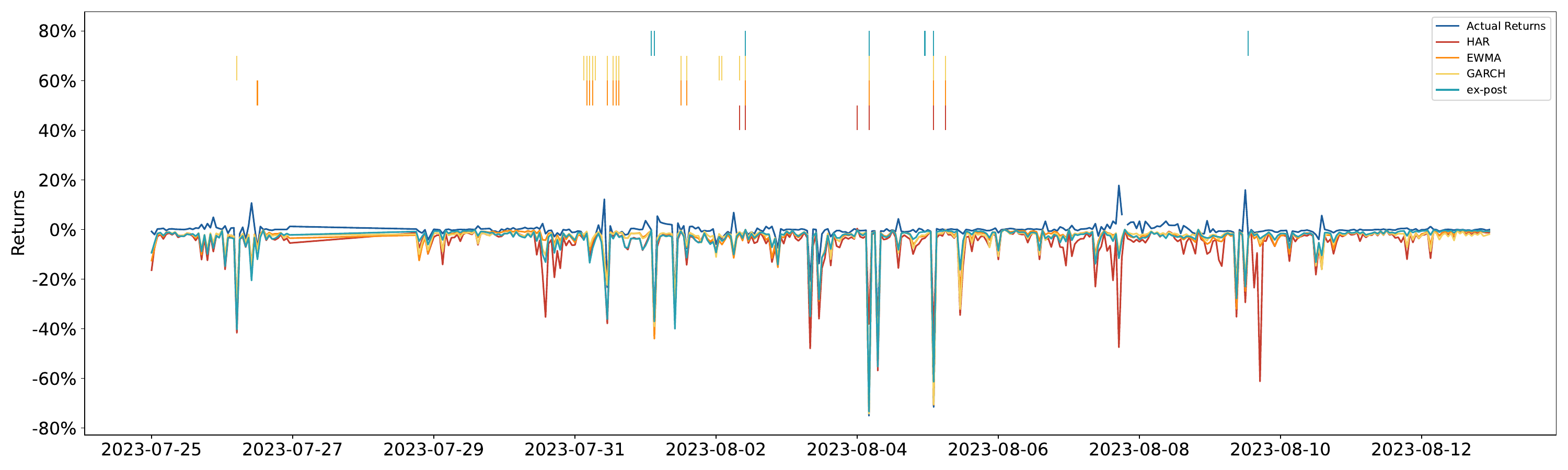}%
}
     \caption{Coverage Test for 1-day VaR}
        \label{fig:coveragetest}
\end{figure}

\subsubsection{Independence Test}
Another crucial aspect to evaluate when validating VaR models is the independence property. In order for VaR forecasts to exhibit accurate conditional coverage, it is essential that past violations do not convey any information about future violations. The failure to meet this criterion can lead to the clustering of violations, which suggests a time lag in the responsiveness of VaR measures to changing market conditions.

In this section of evaluation, we adopted two tests to assess the independence property: Christoffersen's independence test and the multivariate linear regression procedure with F-test.
The former was proposed by Christoffersen as the first approach to test for unconditional coverage and independence separately~\cite{f089d1fa-3082-3934-9893-9bf3432ca0a9}. In this test, relations between VaR violations in successive periods are modelled via a first-order Markow chain. One deficiency in this initial attempt was its limited power in identifying general forms of clustering, as only one lag was considered. The latter belongs to the family of regression-based backtesting procedures which were proposed to overcome the aforementioned limitation by incorporating violations with different lags as well as other information available at the time of forecast. We chose this test because the research by Berger and Moys showed that with small sample sizes, the parsimonious F-test is capable of providing an adequate evaluation of VaR measures~\cite{BERGER2021114893}.

Given that our 413 samples for 1-day VaR estimates were drawn at 1-hour interval over the period of 19 days, it is important to account for the correlation structure in these overlapping observations before applying the independence tests. To mitigate this issue of inherent correlation structure arising from overlapping observations, we separate the samples into distinct groups where samples within each group are drawn from more distant time periods. We acknowledge that the optimal separation strategy is to have 24 groups with each group only contain samples of same timestamp, which will completely mitigate the issue of overlapping observations. Due to the limited number of samples available, such separation will result in multiple groups of no violations for which the F-test cannot be applied to. Here we split the data into 6 groups, with group $i$ contains observations with timestamp at $i$th, $i+6$th, $i+12$th, $i+18$th hours of each day, totaling to nearly 70 samples within each group.

\subsubsection{Christoffersen's independence test}
Following the definition of violation event $i$ in the binomial test, we define $N_{ij}$ as the number of days in which state $j$ occurred in the subsequent period of state $i$. For example, $N_{01}$ represents the number of days for which non-violation is followed by a VaR violation. 

Let $\pi_0$ represents the conditional probability of a violation occurring given there was not violation in the previous period, $\pi_1$ represents the conditional probability of a violation occurring given there was a violation in the previous period. They can be estimated through $\pi_0 = \frac{N_{01}}{N_{00}+N_{01}}$ and $\pi_1 = \frac{N_{11}}{N_{10}+N_{11}}$. Let $\pi$ represents the unconditional probability of a violation occurring, it can be estimated as $\pi = \frac{N_{01}+N_{11}}{N_{00}+N_{01}+N_{10}+N_{11}}$. 

With the null hypothesis for independence as
\begin{equation}
    H_0: \pi_0 = \pi_1 = \pi
\end{equation}
, the test statistic for this test is 
    \begin{equation}
        LR=-2\ln{\frac{(1-\pi)^{N_{00}+N_{10}}(\pi)^{N_{01}+N_{11}}}{(1-\pi_0)^{N_{00}}{\pi_0}^{N_{01}}(1-\pi_1)^{N_{10}}{\pi_1}^{N_{11}}}}
    \end{equation}
This likelihood ratio is asymptotically distributed as a chi-square distribution with 1 degree of freedom, leading to a critical value of 2.706, 3.841 and 6.635 for $10\%$, $5\%$ and $1\%$ significance levels. We present the test statistics and result at $95\%$ confidence level in Table~\ref{tab:independencetest}. 
\begin{table}[htb]
    \scriptsize
    \centering
    \caption{Independence test by group for VaR violations observed with EWMA, DCC-GARCH and HAR estimators. The last column reports the average likelihood ratio weighted by number of samples in each group. }
    \begin{tabular}{cccccccccc}
    \toprule
        \multirow{2}[2]{*}{VaR Level} &\multirow{2}[2]{*}{Estimator} & \multicolumn{7}{c}{Likelihood Ratio}&\multirow{2}[2]{*}{Result} \\
        \cmidrule(lr){3-9}
        {}&{}&Group 0&Group 1&Group 2&Group 3&Group 4&Group 5 & average&{}\\
    \midrule
        \multirow{3}{*}{$95\%$} & {EWMA} &0.28&2.91&0.79&10.44\textsuperscript{***}&1.16&2.88\textsuperscript{*}&3.07\textsuperscript{*}&accept\\
        {} & {GARCH} &0.50&3.29\textsuperscript{*}&0.92&1.25&1.16&1.68&1.47&accept\\
        {} & {HAR} &0.28&0.28&1.71&0.03&0.50&0.12&0.49&accept\\
    \midrule
        \multirow{3}{*}{$97.5\%$} & {EWMA} &0.28&4.91\textsuperscript{***}&0.50&10.44\textsuperscript{***}&0.50&2.88\textsuperscript{*}&3.25\textsuperscript{*}&accept\\
        {} & {GARCH} &0.50&3.28\textsuperscript{*}&0.92&2.12&0.50&1.68&1.50&accept\\
        {} & {HAR} &0.28&0.03&0.12&{N/A}&0.50&{N/A}&0.15&accept\\
    \midrule
        \multirow{3}{*}{$99\%$} & {EWMA} & 0.50 & 0.03& 0.28 & 0.03 & 0.28 & 4.88\textsuperscript{**}&0.99&accept\\
        {} & {GARCH} & 0.28 & 2.91\textsuperscript{*} & 0.92 &0.12& 0.28& 2.88\textsuperscript{*}&1.23&accept\\
        {} & {HAR} & 0.12 & {N/A} & 0.12 & {N/A} & 0.12 & {N/A}&0.06&accept\\
    \bottomrule
    \addlinespace[1ex]
    \multicolumn{8}{l}{\textsuperscript{***}$p<0.01$, 
            \textsuperscript{**}$p<0.05$, 
                \textsuperscript{*}$p<0.1$; N/A where no violations hence no data for LR statistics}
    \end{tabular}    
    \label{tab:independencetest}
\end{table}

At $5\%$ significance level, we observed evidence to reject the independence hypothesis for certain groups in EWMA estimator. Such evidence is statistically significant at all three VaR levels tested, appearing in group 3 for VaR levels of 95\% and 97.5\%, in group 5 for VaR level 99\%. For GARCH and HAR estimator, there was no significant evidence to reject the null hypothesis of independence. When we look at the averaged result, we will not reject any models for failing the independence test at $95\%$ confidence level.
\subsubsection{Regression test}
In this test, violations data were fitted to the multivariate linear model proposed by Christoffersen and Diebold~\cite{ChristoffersenDiebold}:
\begin{equation}
    I_{t} = \alpha + \Sigma_{i=1}^{k}\beta_{1,i}I_{t-i} + \Sigma_{j=1}^{l}\beta_{2,j}g(\cdot) +  u_t
\end{equation}
, where $g(\cdot)$ represents a function on the information set available as of time $t$. In our evaluation, we used $k=4$ and $l=0$. The independence property can be assessed by a simple F-test using the hypothesis
\begin{equation}
    H_0: \beta_{1,1}=\beta_{1,2}=\beta_{1,3}=\beta_{1,4}=0.
\end{equation}

\begin{table}[htb]
    \scriptsize
    \centering
    \caption{F-test statistics by group for the three inference models. The last column reports the weighted average test statistics.} 
    \begin{tabular}{cccccccccc}
    \toprule
        \multirow{2}[2]{*}{VaR Level} &\multirow{2}[2]{*}{Estimator} & \multicolumn{7}{c}{Test statistics}&\multirow{2}[2]{*}{Result} \\
        \cmidrule(lr){3-9}
        {}&{}&Group 0&Group 1&Group 2&Group 3&Group 4&Group 5& average&{}\\
    \midrule
        \multirow{3}{*}{$95\%$} & {EWMA} & 0.17 & 1.90 & 0.80 & 17.30\textsuperscript{***} & 1.27 & 3.13\textsuperscript{**} & 4.10\textsuperscript{***}&reject\\
        {} & {GARCH} & 0.78 & 1.83 & 0.78 & 1.69 & 0.41 & 1.19 &1.11&accept\\
        {} & {HAR} & 0.17 & 0.17 & 0.87 & 0.02 & 0.78 & 0.07 &0.35&accept\\
    \midrule
        \multirow{3}{*}{$97.5\%$} & {EWMA} & 0.17 & 9.56\textsuperscript{***} & 0.78 & 17.30\textsuperscript{***} & 0.36 & 3.31\textsuperscript{**}&5.25\textsuperscript{***}&reject\\
        {} & {GARCH} & 0.78 & 1.83 & 0.78 & 2.78\textsuperscript{**} & 0.78 & 1.19&1.35&accept\\
        {} & {HAR} & 0.17 & 0.07 & 0.07 & {N/A} & 0.78 & {N/A}&0.27&accept\\
    \midrule
        \multirow{3}{*}{$99\%$} & {EWMA} & 0.34 & 0.02 & 1.54 & 0.02 & 0.17 & 9.39\textsuperscript{***}&1.90&accept\\
        {} & {GARCH} & 0.17 &1.90 & 0.78 & 7.18\textsuperscript{***} & 0.17 & 3.13\textsuperscript{**}&2.21&accept\\
        {} & {HAR} & 0.07 & {N/A} & 0.07 & {N/A} & 0.07 & {N/A}&0.07&accept\\
    \bottomrule
    \addlinespace[1ex]
    \multicolumn{8}{l}{\textsuperscript{***}$p<0.01$, 
            \textsuperscript{**}$p<0.05$, 
                \textsuperscript{*}$p<0.1$; N/A where no violations hence no data for F-test}
    \end{tabular}    
    \label{tab:ftest}
\end{table}

In Table~\ref{tab:ftest} we recorded the test statistics for each group at different VaR levels and with different volatility estimators. In contrast to the conclusion from the previous independence test, we observed statistically significant evidence to reject the independence property for EWMA estimator for VaR levels of $95\%$ and $97.5\%$. For GARCH estimator, F-test applied to group 3 and group 5 found evidence to reject the null hypothesis for $97.5\%$ and $99\%$ VaR. But this evidence becomes insignificant when we assessed it at the averaged level. For HAR estimator, there was no statistically significant p-value to reject the null hypothesis of Independence. We noted the per-estimator and per-group result from F-test were reasonably consistent with the result from the previous independence test.

\subsection{Summary of Accuracy Performance}
We summarise the results of the backtests applied to assess the unconditional coverage and independence properties of VaR forecast in Table~\ref{tab:backtest} below. Overall, HAR has the best performance in terms of accuracy.

\begin{table}[htb]
    \small
    \centering
    \caption{Hit rates(i.e.,~the fraction of violations) for the three inference models. The column Passes shows the total number of times the model passes any of the three tests. Best possible score is 9/9 (3 VaR-levels times 3 tests). For independence tests where tests were conducted to multiple groups, the pass/fail is decided by the average of the test statistics across 6 groups.} 
    \begin{tabular}{ccccc}
    \toprule
        \multirow{2}{*}{Inference model} & \multicolumn{3}{c}{VaR Level} & \multirow{2}{*}{Passes} \\
        \cmidrule(lr){2-4}
        {}&$95\%$&$97.5\%$&$99\%$& \\
    \midrule
        {EWMA} & 7.56\% & 6.18\% & 4.58\% & 5/9\\
        {GARCH} & 9.84\% & 9.15\% & 6.64\% & 6/9\\
        {HAR} & 4.58\% & 3.66\% & 3.66\% & 9/9\\
    \bottomrule
    \end{tabular}    
    \label{tab:backtest}
\end{table}

%%%%%%%%%%%%%%%%%%%%%%%%%%%%%%%%%%%%
\section{Conclusion \& Future Work}\label{chap:Conclusion}

\subsection{Summary}
In this work, we have presented a real-time VaR calculation workflow for portfolios of cryptocurrency derivatives. 

From the perspective of workflow design, we applied a parsimonious volatility forecast model which can be fitted with OLS estimators to ensure computation efficiency. To further reduce the calculation latency, delta-gamma-theta approach is chosen as a replacement of the commonly used historical or Monte Carlo simulation approach, to approximate the non-Gaussian distribution of portfolio returns. As the final step in this computation workflow, we applied the Cornish-Fisher expansion to enhance the estimation for tail quantiles of the distribution with higher order moments. 

From the perspective of implementation, we leveraged the column-oriented approach of kdb+ database and in-memory compute engines to ensure efficiency in dealing with high-frequency market data. We developed a customised kdb+ tick architecture that pre-processes tick level data to facilitate the calculation workflow. For the calculation process, we included tables for caching and latest market data to decrease the latency from IPC. Where effective, parallel processing is used to further enhance the calculation efficiency. To improve the usability of the system, we developed a complementary web-based interface in KX dashboard, which provide users with access VaR calculation within a few clicks and other relevant market metrics to assist the risk management practice. 

As part of this work, we also conducted a comparative analysis involving three distinct families of volatility models: EWMA, GARCH and HAR. These inference models were evaluated based on two critical dimensions: calculation latency and VaR estimation accuracy. While EWMA exhibited superior performance in terms of calculation latency due to its computational simplicity, it fell short in adequately capturing the dynamics of volatility, resulting in clustered violations and underestimations of the risk level. In contrast, the HAR model emerged as the top performer in terms of inference accuracy by successfully passing the backtests on unconditional coverage and independence.

\subsection{Future Work}

\subsubsection{Positive semi-definiteness of covariance matrix}
In the context of our inference algorithm, the elements within the covariance matrix are forecasted independently. We have observed instances in which the forecasted correlation terms deviate from the valid range of -1 to 1, resultin in a covariance matrix that is not positive semi-definite. This statistically incoherent covariance matrix leads to negative values for second order central moments, consequently rendering the skewness parameter unsuitable for the Cornish-Fisher expansion.

One potential correction procedure involves projecting the symmetric matrix onto the space of positive semi-definite matrices, as illustrated in the study by Fan et al~\cite{Fan}. Other systematic remedies have been proposed by researchers. For instance, the reparameterisation approach introduced by Archakov and Hansen transforms the $n \times n$ correlations matrix into unrestricted vectors of length $n(n-1)/2$ for modelling correlations. This ensures the correlations matrix forecast obtained through inverse mapping retains the intrinsic property of positive definiteness~\cite{Archakov}.

\subsubsection{High dimensional covariance matrix}
The challenge from forecasting high dimensional covariance matrix is twofolds: firstly, number of covariance or correlation terms increases quadratically as number of underlyings increases, resulting in issue on computation efficiency; secondly, considering the growth in the number of digital assets the number of samples available could be less than the number of covariance terms to forecast.

\subsubsection{Validity of Cornish-Fisher expansion}
While Cornish-Fisher expansion provides a relatively easy and parsimonious way of dealing with non-normal return distributions, its usefulness may be compromised by two pitfalls as discussed in the work of Maillard~\cite{Maillard}:
\begin{itemize}
    \item \textbf{Domain of validity}: For Cornish-Fisher expansion to produce a well-defined monotonic quantile function, it requires the actual skewness $S$ and excess kurtosis $K$ of the return distribution to satisfy conditions of~\cite{Amédée-Manesme2015}:
\begin{equation}
    \frac{S^2}{9} - 4 (\frac{K}{8} - \frac{S^2}{6})(1-\frac{K}{8}-\frac{5S^2}{36}) \leq 0
\end{equation}
When $S$ and $K$ fail outside of this domain, the Cornish-Fisher expansion is no longer applicable. We can consider applying the rearrangement procedure introduced by Chernozhukov et al to restore the monotonic property inherent in the quantile functions~\cite{8d0c0059-3b65-384a-981e-2ed07a73d4da}.

\item\textbf{Skewness and Kurtosis parameter}: 
It is important to differentiate between the actual skewness $S$ and excess kurtosis $K$ of the true distribution and the $\hat{S}$ and $\hat{K}$ parameters applied in the transformation, as they only coincide when their values are small~\cite{Maillard}. Maillard presented actual skewness and excess kurtosis as polynomials of $\hat{S}$ and $\hat{K}$ parameters:
\begin{align}
    S &= f(\hat{S}, \hat{K}) = \frac{6S-76S^3+510S^5+36SK-468S^3K+108SK^2}{(1+6K^2-24S^2K+25S^4)^{1.5}}\\
    K &= g(\hat{S}, \hat{K}) = \frac{\splitdfrac{\splitdfrac{3+3348K^4-28080S^2K^3+1296K^3+252K^2+24K}%
                          {-123720S^6K + 8136S^4K - 504S^2K-6048S^2K^2}}%
                          {+64995S^8-2400S^6-42S^4+88380K^2S^4}}{(1+6K^2-24S^2K+25S^4)^{2}} - 3
\end{align}
    
and demonstrated that to appropriately apply the Cornish-Fisher expansion, it is necessary to reverse these polynomials to solve for $\hat{S}$ and $\hat{K}$. While analytical expressions for the reversed relations are not available, to improve the quality of transformation, we could use numeric solvers for these polynomials, such as the modified Newton's method proposed in the work of Lamb et al~\cite{2de53256144c4962bbdecf0d8bc16c42}.

\end{itemize}

%% bibliography
\bibliographystyle{simple}
%\bibliography{export}

\end{document}